\documentstyle{article}
\input{ epsf }
\begin{document}
\title{$k$-Gerbes, Line Bundles and Anomalies}
\author{{\bf C. Ekstrand}\\Department of Theoretical Physics, \\Royal Institute of
Technology, \\S-100 44 Stockholm, Sweden \\ ce@theophys.kth.se 
}
\date{}
\maketitle
\newcommand{\eq}{\begin{equation}}
\newcommand{\eqend}{\end{equation}}
\newcommand{\eqa}{\begin{eqnarray}}
\newcommand{\eqaend}{\end{eqnarray}}
\newcommand{\nonu}{\nonumber \\ \nopagebreak}
\newcommand{\Ref}[1]{(\ref{#1})}
\newcommand{\B}{{\cal B}}
\newcommand{\bbar}{\bar}
\newcommand{\F}{{\cal F}}
\newcommand{\cL}{{\cal L}}
\newcommand{\ep}{{\cal E}}
\newcommand{\EE}{{\mbox{\scriptsize\bf E}}}
\newcommand{\E}{{\bf E}}
\newcommand{\SC}{{\bf{A}}}
\newcommand{\CU}{{\bf{F}}}
\newcommand{\W}{{\cal W}}
\newcommand{\M}{\cal M}
\newcommand{\Ss}{S}
\newcommand{\A}{{\cal A}}
\newcommand{\T}{\cal T}
\newcommand{\D}{\cal D}
\newcommand{\g}{\cal g}
\newcommand{\G}{{\cal G}}
\newcommand{\Rr}{\cal R}
\newcommand{\Pp}{\cal P}
\newcommand{\etta}{X}
\newcommand{\C}{\cal C}
\newcommand{\V}{\cal V}
\newcommand{\mmu}{\mu}
\newcommand{\mmmu}{\lambda ^{\prime}}
\newcommand{\mdim}{{2n}}
\newcommand{\ddd}{d}
\newcommand{\uu}{u}
\newcommand{\ommega}{\omega ^{\prime}}
\newcommand{\ommmega}{\omega }
\newcommand{\nejnum}{\nonumber}
\def\eop{\nopagebreak\hfill $\Box$}
\newtheorem{definition}{Definition}
\newtheorem{lemma}{Lemma}
\newtheorem{theorem}{Theorem}
\newtheorem{proposition}{Proposition}
\newtheorem{corollary}{Corollary}
\begin{abstract}
 We use sets of trivial line bundles for the realization of gerbes. For $1$-gerbes the structure arises naturally for the Weyl fermion vacuum bundle at a fixed time. The Schwinger term is an obstruction in the triviality of a 1-gerbe.
\end{abstract}
\section{Introduction}
Gerbes were introduced by Giraud \cite{GI} in the early seventies. However, their importance in physics were not appreciated until about twenty years later. Then publications by authors as Brylinski and Freed appeared, see for instance \cite{BR,F}. Since then, the number of people working on gerbes has increased for each year. Their importance in string theory will probably lead to that this development will continue. The drawback is that the theory behind gerbes contains mathematics which is unfamiliar for most physicists. Even the pure definition of a gerbe, as a sheaf of groupoids with certain properties, is not easy. This initiated the authors of \cite{TH,CH,MU} to construct simple differential geometrical objects for the realization of gerbes. In these papers, and in the present, the following loose definition is used: A $(k-1)$-gerbe with band ${\bf C}^\times $ on a manifold $B$ is a set of elements (which preferable can be given some geometrical meaning) with an equivalence relation so that the set of equivalence classes can be given a group structure so it is isomorphic with the \v{C}ech cohomology $\check{H}^k(B,\underline{{\bf C}}^\times )$, $k\geq 2$. Certainly, this is not the same as the original definition, but the 1-gerbes in \cite{TH,CH,MU} are actually also gerbes in Giraud's sense. Further, our definition agrees with what we want for $k=1$, i.e. that line bundles are classified by $\check{H}^1(B,\underline{{\bf C}}^\times )$. 

In this paper, we will review and develop a construction that was attempted in \cite{TH}, where certain sets of local (trivial) line bundles are interpreted as $k$-gerbes. Independently, this construction has also been made by Hitchin and Chatterjee for the case of 1-gerbes, \cite{CH}. Closely related are also the bundle gerbes used by Murray \cite{MU}. Since it appears as some papers misinterpret the construction in \cite{TH,CH} we find it worth to develope it in detail. 

To motivate the importance of sets of local line bundles we will show how they appear naturally from the vacuum line bundle. A geometrical description of the vacuum and the Schwinger term in terms of gerbes has previously been made in \cite{CMM} and in many other articles based on \cite{MU}. The constructions in these papers are however complicated and artificial due to two reasons: First, the bundle gerbes do not allow any nice generalization to higher order gerbes. Secondly, to obtain an isomorphism with $\check{H}^2(B,\underline{{\bf C}}^\times )$ one have to use equivalence of bundle gerbes in a sense similar to what is made in K-theory. 

 We will derive the Schwinger term from the curvature of the determinant line bundle for a manifold with boundary. The derivation is close to the one made in \cite{CMM}.
 We come to the conclusion that the Schwinger term, with the relevant cohomology, can be interpreted as an obstruction in a set of local line bundles on $\A /\G$ which are equipped with a certain structure. This is the analog of the geometrical description of the chiral anomaly in the space-time formalism given by Atiyah and Singer \cite{AS}. This shows that the description of gerbes as local line bundles is no abstract nonsense and in fact appears in a natural way, at least for 1-gerbes.

 We will start with proving that gerbes can be described by sets of local line bundles with a certain structure. This will be done in section 2. In section 3 we review the \v{C}eck-de Rham cohomology and in section 4 we pull back a part of it to an affine space so that new interesting cohomology groups are obtained. Using the mathematical results of section 3 and 4 we show in section 5 that the geometrical structure which has been developed appears naturally in the context of anomalies and Schwinger terms.

\section{\v{C}ech cohomology and sets of line bundles}
 We will study the \v{C}ech cohomology with respect to an open covering $\{ U_\lambda \} _{\lambda\in \Lambda}$ of a manifold $B$. We don't put any restriction on the dimension of $B$, but we assume that it (and all other manifold which we will discuss) is smooth and Hausdorff. Though not necessary, we will for pedagogical reasons assume that the index set $\Lambda$ is a chain, or equivalently, a linearly ordered set. 
\begin{definition}
\label{d:1}
 A \v{C}ech cochain $\{ g_{\lambda _0 ... \lambda _k}\} \in \check{C}^k(\{ U_\lambda \} ,\underline{{\bf C}}^\times )$ is a set consisting of 
smooth functions $ g_{\lambda _0 ... \lambda _k}:U_{\lambda _0 ... \lambda _k}\rightarrow {\bf C}^\times ={\bf C}\setminus \{ 0\} $ for every ordered ensemble $\lambda _0<...<\lambda _k$ such that the intersection $U_{\lambda _0 ... \lambda _k}=U_{\lambda _0}\cap ... \cap U_{\lambda _k}$ is non-empty. The set $\check{C}^k(\{ U_\lambda \},\underline{{\bf C}}^\times )$ of cochains is given the structure of an abelian group according to 
$(g\cdot g^\prime )_{\lambda _0 ... \lambda _k}(b)=g_{\lambda _0 ... \lambda _k} (b) g^\prime _{\lambda _0 ... \lambda _k}(b)$, $(g^{-1})_{\lambda _0 ... \lambda _k}(b)=g_{\lambda _0 ... \lambda _k}(b)^{-1}$ and ${\bf 1}_{\lambda _0 ... \lambda _k}(b)=1\in {\bf C}^\times $, for $b\in U_{\lambda _0 ... \lambda _k}$. Let $\delta :\check{C}^k(\{ U_\lambda \} ,\underline{{\bf C}}^\times )\rightarrow \check{C}^{k+1}(\{ U_\lambda \} ,\underline{{\bf C}}^\times )$ be the homomorphism given by 
\eq
\label{eq:CH}
(\delta \{ g_{\mu _0 ... \mu _k}\} )_{\lambda _0 ... \lambda _{k+1}}= \prod _{i=0}^{k+1}(g_{\lambda _0 ... \hat{\lambda }_i ...\lambda _{k+1}}|_{U_{\lambda _0 ... \lambda _{k+1}}})^{(-1)^i}.
\eqend
The group $\check{H}^k(\{ U_\lambda \} ,\underline{{\bf C}}^\times )=\mbox{ker}\delta \cap \check{C}^k(\{ U_\lambda \} ,\underline{{\bf C}}^\times ) / \mbox{im}\delta \cap \check{C}^k(\{ U_\lambda \} ,\underline{{\bf C}}^\times )$ is called the \v{C}ech cohomology of degree $k$ with respect to the covering $\{ U_\lambda \}$ of $B$.
\end{definition}
 It is well-known that (up to isomorphisms) the \v{C}ech cohomology is independent of the covering as long as one restricts to good (or Leray) coverings, i.e. such that all non-empty intersections $U_{\lambda _0 ... \lambda _k}$ are contractible. This group is called the \v{C}ech cohomology $\check{H}^k(B,\underline{{\bf C}}^\times )$ of $B$.  Well studied is $\check{H}^1(B,\underline{{\bf C}}^\times )$ since it classifies line bundles, a fact which we now will review. Unless stated otherwise, it will for the rest of the paper be assumed that the covering is open and good and that $B$ admits such coverings. The existence of such coverings is clear when $B$ is of finite dimension, see for instance \cite{BT}. 

 Let $\cL$ be a line bundle on $B$.
 Decompose it into trivial pieces $\cL |_{U_\lambda }$, choose non-vanishing sections $s_\lambda$ and define $g_{\lambda \mu }$ by
\eq
\label{eq:SSG}
s_\lambda =s_\mu g_{\lambda \mu } \quad \mbox{on } U_{\lambda \mu }.
\eqend
 This equation is actually not well-defined. The reason is that $s_\lambda $ is a section of $\cL |_{U_\lambda }$ while $s_\mu g_{\lambda \mu } $ is a section of $\cL |_{U_\mu }$. We have an equality of sections belonging to different line bundles. The correct way to make sense of this is to introduce an isomorphism $\chi _{\lambda \mu }:\cL |_{U_\lambda }\rightarrow\cL |_{U_\mu }$ and use it for identification. Then eq. \Ref{eq:SSG} becomes:
\eq
\label{eq:SGS}
\chi _{\lambda \mu } (s_\lambda )=s_\mu g_{\lambda \mu } \quad \mbox{on } U_{\lambda \mu }.
\eqend
This is however implicit in eq. \Ref{eq:SSG} where $\chi _{\lambda \mu }$ has been chosen as the canonical isomorphism which is obtained from the fact that  $\cL |_{U_\lambda }$ and $\cL |_{U_\mu }$ are restrictions of a global bundle $\cL$. From a line bundle we have thereby constructed a triple $(\cL _\lambda ,s_{\lambda }, \chi _{\lambda \mu })$, where $\chi _{\lambda \mu }$ in this case is a canonical isomorphism. The set $\{ g_{\lambda \mu }\}$, defined by eq. \Ref{eq:SGS}, represents an element in $\check{H}^1(B,\underline{{\bf C}}^\times )$ that is independent of the choice of $s_\lambda $'s, the covering and the choice of representative of the element $[\cL ]$ in the equivalence class of line bundles.

 Conversely, let a \v{C}ech cocycle $\{ g_{\lambda \mu }\}$ be given and choose line bundles $\cL _\lambda \rightarrow U_\lambda $ with sections $s_\lambda $ (the choices $\cL _\lambda = U_\lambda \times {\bf C}$ and $s_\lambda =1_\lambda $ are frequent in the literature). Then define $\chi _{\lambda \mu}$ by eq. \Ref{eq:SGS}. Again, we have constructed a triple $(\cL _\lambda ,s_\lambda ,\chi _{\lambda \mu})$. Eq. \Ref{eq:SGS} can be regarded as an equivalence relation in the set $\{ \cL _\lambda \}$. By taking the quotent we obtain a smooth line bundle whose equivalence class is independent of the choices that were made. 

 The two maps that were described above are homomorphisms and they are 
each others inverses. We have thus established an isomorphism between $\check{H}^1(B,\underline{{\bf C}}^\times )$ and the equivalence class of line bundles. As a \lq middle step\rq $ $ of each homomorphism, a triple appears. It is clear that if a certain equivalence relation and group structure is introduced on triples, we obtain a new group which is isomorphic to $\check{H}^1(B,\underline{{\bf C}}^\times )$. That the class of $\{ g_{\lambda \mu }\}$ is independent of $\{ s_\lambda \}$ means that it is only equivalence classes of pairs $(\cL _\lambda ,\chi _{\lambda \mu})$ that are important from a cohomological viewpoint. We have thereby established:
\eq
\label{eq:TEXT}
\mbox{\v{C}ech cohomology}\cong\{\mbox{Equivalence classes of pairs}\}\cong\{\mbox{Classes of line bundles}\} 
\eqend
We will soon show how parts of this can be generalized to $\check{H}^k(B,\underline{{\bf C}}^\times )$. First, we will however reinterpret $\chi _{\lambda \mu }$. This may appear strange at first, but in the generalization to arbitrary values of $k$ the idea will become clear. Introduce for every $\lambda $ an isomorphism $\xi _\lambda $ from $\cL _\lambda $ to its dual $\cL _\lambda ^{-1}$, i.e. $\xi _\lambda (s_\lambda )=s_\lambda ^{-1}$ for some section $s_\lambda ^{-1}$ on $\cL _\lambda ^{-1}$ and $\xi _\lambda (s_\lambda r_\lambda )=s_\lambda ^{-1}/r_\lambda $, where $r_\lambda :U_\lambda \rightarrow {\bf C}^\times$. Then $\chi _{\lambda \mu }$ defines a section $\sigma _{\lambda \mu }$ on $\cL _\lambda \otimes\cL _\mu ^{-1}$ according to: $\sigma _{\lambda \mu }=s_\lambda \otimes \xi _\mu (\chi _{\lambda \mu }(s_\lambda ))$. Clearly, $\sigma _{\lambda \mu }$ is independent of the choice of $s_\lambda $. It defines a bijective correspondence between $\chi _{\lambda \mu }$ and pairs $(\sigma _{\lambda \mu },\xi _\mu )$ subject to the equivalence relation $(\sigma _{\lambda \mu }r_\mu,r_\mu\xi _\mu )\sim (\sigma _{\lambda \mu },\xi _\mu )$. Thus, for a fixed choice of $\{ \xi _\lambda \} $, isomorphisms $\chi _{\lambda \mu }$ can be identified with sections $\sigma _{\lambda \mu }$. 

 We regard the above discussion to be motivation enough for generalization to $\check{H}^k(B,\underline{{\bf C}}^\times )$. 
 Let $\{ \cL _{\lambda _0 ...\lambda _{k-1}}\}$ and $\{ s _{\lambda _0 ...\lambda _{k-1}}\}$ be (trivial) line bundles with corresponding 
non-vanishing sections over all $U_{\lambda _0 ...\lambda _{k-1}}$ (non-empty sets and with ordered indices as usual). Let $\{ \xi _{\lambda _0 ...\lambda _{k-1}}\}$ be a fixed choice of isomorphisms from $\cL _{\lambda _0 ...\lambda _{k-1}}$ to its dual. For instance, when $\cL _{\lambda _0 ...\lambda _{k-1}}$ admits a non-degenerate metric we can choose $\{ \xi _{\lambda _0 ...\lambda _{k-1}}\}$ and the dual bundle as the ones induced by the metric. Notice that in the construction that follows we will avoid discussing equality between the line bundles themselves, because all line bundles over a contractible set $U_{\lambda _0 ...\lambda _{k-1}}$ are isomorphic. We thus need a different concept of equality for sets of line bundles. This is definition \Ref{d:3} below. We define $\{ (\delta\cL )_{\lambda _0 ...\lambda _{k}}\}$ and $\{ (\delta s )_{\lambda _0 ...\lambda _{k}}\}$ over $U_{\lambda _0 ...\lambda _{k}}$ in the spirit of eq. \Ref{eq:CH}:
\eqa
\label{eq:LS}
 (\delta\cL )_{\lambda _0 ...\lambda _{k}} & := &\cL _{\lambda _1\lambda _2 ...\lambda _{k}}\otimes {\cL _{\lambda _0\lambda _2 ...\lambda _{k}}}^{-1}\otimes ...\otimes 
{\cL _{\lambda _0\lambda _1 ...\lambda _{k-1}}}^{(-1)^{k}}\nonu
(\delta s )_{\lambda _0 ...\lambda _{k}}& := & s_{\lambda _1\lambda _2 ...\lambda _{k}}\otimes {s_{\lambda _0\lambda _2 ...\lambda _{k}}}^{-1}\otimes ...\otimes {s_{\lambda _0\lambda _1 ...\lambda _{k-1}}}^{(-1) ^{k}},
\eqaend
where $a\otimes b\otimes c\otimes ...:=(((a\otimes b)\otimes c)\otimes ...)$ and ${s_{\lambda _0 ...\lambda _{k-1}}}^{-1}:=\xi _{\lambda _0 ...\lambda _{k-1}}(s_{\lambda _0 ...\lambda _{k-1}})$. It is understood that the line bundles and sections on the right hand side are restricted to $U_{\lambda _0 ...\lambda _{k}}$. 
 Since $(\delta\cL ^{-1})_{\lambda _0 ...\lambda _k}$ can be identified with the dual of $(\delta\cL )_{\lambda _0 ...\lambda _k}$, it is clear how $\xi _{\lambda _0...\lambda _{k-1}}$ induces an isomorphism $(\delta \xi )_{\lambda _0...\lambda _{k}}$ from $(\delta\cL )_{\lambda _0 ...\lambda _k}$ to its dual. Notice that there exist a canonical section $(\delta ^2s)_{\lambda _0...\lambda _{k+1}}$ on $(\delta ^2\cL )_{\lambda _0...\lambda _{k+1}}$. The section is independent of the choice of $\{ s_{\lambda _0...\lambda _{k-1}}\}$. This makes it possible to canonically identify $(\delta ^2\cL )_{\lambda _0...\lambda _{k+1}}$ with the product bundle $U_{\lambda _0...\lambda _{k+1}}\times {\bf C}$ (which have a canonical section taking the value $1\in {\bf C}$). We will therefore think of line bundles of the type $(\delta ^2\cL )_{\lambda _0...\lambda _{k+1}}$ as an identity. We will then consider the cohomology generated when $\delta $ acts on sets $\{ ( \cL ^{\prime }_{\lambda _0 ...\lambda _k},s^{\prime }_{\lambda _0 ...\lambda _k})\}$ according to \Ref{eq:LS}. Thus, the cocycles should be such that $\delta $ acting on them will give the identity $\{ (\delta ^2\cL )_{\lambda _0...\lambda _{k+1}},(\delta ^2s)_{\lambda _0...\lambda _{k+1}} \}$, for some set $\{ ( \cL _{\lambda _0 ...\lambda _{k-1}},s_{\lambda _0 ...\lambda _{k-1}})\}$. It implies that $\cL ^{\prime }_{\lambda _0 ...\lambda _k}:=(\delta\cL )_{\lambda _0 ...\lambda _k}$ and $(\delta s ^{\prime })_{\lambda _0 ...\lambda _{k+1}}=(\delta ^2s)_{\lambda _0...\lambda _{k+1}}$. We have thereby motivated the following definition:
\begin{definition}
\label{d:2}
 A set $\{ (\cL _{\lambda _0 ...\lambda _{k-1}}, \sigma _{\lambda _0 ...\lambda _{k}})\}$ of pairs consists of line bundles $\cL _{\lambda _0 ...\lambda _{k-1}}$ (with isomorphisms $\xi _{\lambda _0 ...\lambda _{k-1}}$ to the duals) on each $U_{\lambda _0 ...\lambda _{k-1}}$ and non-vanishing sections $\sigma _{\lambda _0 ...\lambda _{k}}$ on each $(\delta \cL )_{\lambda _0 ...\lambda _{k}}$ such that $(\delta \sigma )_{\lambda _0 ...\lambda _{k+1}}=(\delta ^2s)_{\lambda _0...\lambda _{k+1}}$. 
\end{definition}
\begin{definition}
\label{d:3}
$\{ (\cL _{\lambda _0 ...\lambda _{k-1}}, \sigma _{\lambda _0 ...\lambda _{k}})\}$ and $\{ (\cL ^\prime _{\lambda _0 ...\lambda _{k-1}}, \sigma ^\prime _{\lambda _0 ...\lambda _{k}})\}$ are said to be equivalent if there exist a diffeomorphism between them, i.e. diffeomorphisms $\varphi _{\lambda _0 ...\lambda _{k-1}}: \cL _{\lambda _0 ...\lambda _{k-1}}\rightarrow \cL ^\prime _{\lambda _0 ...\lambda _{k-1}}$ such that $(\delta \varphi )_{\lambda _0 ...\lambda _{k}}(\sigma _{\lambda _0 ...\lambda _{k}})=\sigma ^\prime _{\lambda _0 ...\lambda _{k}}$.
\end{definition}
 Obviously, $(\delta \varphi )_{\lambda _0 ...\lambda _{k}}$ is defined in accordance with eq. \Ref{eq:CH} and $\varphi _{\lambda _0 ...\lambda _{k-1}}^{-1}:=\xi _{\lambda _0 ...\lambda _{k-1}}^{\prime }\circ\varphi _{\lambda _0 ...\lambda _{k-1}}\circ\xi _{\lambda _0 ...\lambda _{k-1}}^{-1}$. 
 Two sets of pairs can be multiplied to give $\{ (\cL _{\lambda _0 ...\lambda _{k-1}}\otimes \cL ^\prime _{\lambda _0 ...\lambda _{k-1}}, (\sigma \otimes\sigma ^\prime )_{\lambda _0 ...\lambda _{k}})\}$, where $(\sigma \otimes\sigma ^\prime )_{\lambda _0 ...\lambda _{k}}$ is the image of $\sigma _{\lambda _0 ...\lambda _{k}}\otimes\sigma ^\prime _{\lambda _0 ...\lambda _{k}}$ under the canonical isomorphism from $(\delta \cL )_{\lambda _0 ...\lambda _{k}}\otimes (\delta \cL ^{\prime })_{\lambda _0 ...\lambda _{k}}$ to $(\delta (\cL \otimes \cL ^{\prime }))_{\lambda _0 ...\lambda _{k}}$. This defines an abelian group $H^k(\cL ,\sigma )$ whose elements are the equivalence classes of (sets of) pairs. The identity element consist of pairs $(\cL _{\lambda _0 ...\lambda _{k-1}}, \sigma _{\lambda _0 ...\lambda _{k}})$ such that there exist sections $s_{\lambda _0 ...\lambda _{k-1}}$ with $(\delta s) _{\lambda _0 ...\lambda _{k}}=\sigma _{\lambda _0 ...\lambda _{k}}$.
\begin{proposition}
\label{p:1}
$H^k(\cL ,\sigma )\cong\check{H}^k(B,\underline{{\bf C}}^\times )$.
\end{proposition}
\vspace{2mm}
{\bf Proof}\hspace{2mm}
 It follows directly from the equation 
\eq
\label{eq:BI}
(\delta s)_{\lambda _0 ...\lambda _{k}}=\sigma _{\lambda _0 ...\lambda _{k}}g_{\lambda _0 ...\lambda _{k}},
\eqend
where $s_{\lambda _0 ...\lambda _{k-1}}$ is a non-vanishing section of $\cL _{\lambda _0 ...\lambda _{k-1}}$.
\eop

 This implies that $H^k(\cL ,\sigma )$ is independent of the covering, as indicated by the notation. 
 We have thus verified the left isomorphism in \Ref{eq:TEXT} for all $k\geq 1$. Unfortunately, we have not accomplished to establish an equivalence of the isomorphism to the right.
 However, this was not to expect since we started of with line bundles which obviously are the wrong geometrical objects for this. We just have to accept that we can only come half-way in \Ref{eq:TEXT} and instead try to make the best out of it. 

 Let us consider the case $k=0$. Then a pair $\{(\cL , \sigma _\lambda )\}$ consist  by definition of a trivial line bundle $\cL$ over $B$ and non-vanishing sections $\sigma _\lambda$ 
of $\cL _\lambda =(\delta\cL )_\lambda =\cL |_{U_\lambda }$. A pair is a cocycle if $(\delta \sigma )_{\lambda \mu }=(\delta ^2s)_{\lambda\mu }$. Notice that $(\delta ^2s)_{\lambda\mu }$
 is the canonical section of $\cL _\mu \otimes\cL_\lambda ^{-1}$ given by $s|_{U_\mu } \otimes (s|_{U_\lambda })^{-1}$. Thus, the cocycle relation means that the $\sigma$'s glue together to define a global section. The corresponding \v{C}ech cocycle, given by $s|_{U_\lambda }=\sigma _\lambda g_\lambda ,$ then also glue together to define a smooth ${\bf C}^\times$-valued function on $B$. Thus, we see that elements in $\check{H}^0(B,\underline{{\bf C}}^\times )$, which are ${\bf C}^\times$-valued function on $B$, can be identified with the \lq quotient\rq $ $ of two non-vanishing sections of a trivial bundle on $B$. Notice that the case $k=0$ is special in our setting: it is necessary to introduced a fix choice of section $s$ on $\cL$ (because $g$ depends on $s$). 

We have already considered $k=1$, so lets turn to the complicated case when $k\geq 2$. One way to get information about how (sets of) pairs look in these cases is to study the case when $B$ is contractible.
 Then everything becomes trivial so a degree $k$ cocycle can be written as the coboundary operator acting on a $k-1$ cochain, which often is simpler to understand. Consider for example the case $k=1$.
 Then triviality means that there exist a representative $\{(\cL _\lambda ,\sigma _{\lambda \mu })\}$ of a cohomology class of pairs which is of the form $\delta \{(\cL ^\prime ,\sigma _{\lambda }^\prime )\}$.
 That $\cL _\lambda \equiv (\delta\cL ^\prime )_\lambda$ means that the $\cL _\lambda$'s are restrictions of a global trivial bundle $\cL ^\prime$. That a line bundle over a contractible space is a trivial line bundle we already knew, but here the setting of pairs gave us the answer. When $k=2$ triviality means that a representative $\{ (\cL _{\lambda\mu } ,\sigma _{\lambda \mu \nu})\}$ of a cohomology class of pairs is of the form $\delta \{ (\cL ^\prime _\lambda ,\sigma _{\lambda \mu}^\prime)\}$.
 Thus, in this case we get: $\cL _{\lambda\mu }\equiv\cL ^\prime _{\mu }\otimes {\cL ^\prime _{\lambda }}^{-1}$ and $\sigma _{\lambda\mu\nu }\equiv\sigma ^\prime _{\mu\nu }\otimes {\sigma ^\prime _{\lambda\nu }}^{-1}\otimes \sigma ^\prime _{\lambda\mu }$. Observe that 
 $\{(\cL ^\prime _\lambda ,\sigma _{\lambda \mu}^\prime)\}$ does not describe a global line bundle since it is not a cocycle. The \lq transition functions\rq $ $ do not obey the cocycle relation and the local lines do therefore not glue together.   

 The pairs $(\cL _{\lambda _0 ...\lambda _{k-1}}, \sigma _{\lambda _0 ...\lambda _{k}})$ are actually independent of the first argument $\cL _{\lambda _0 ...\lambda _{k-1}}$. The line bundles are only needed to define the $\sigma _{\lambda _0 ...\lambda _{k}}$'s and can thus be chosen to be product bundles. Thus, the \v{C}ech cohomology can be interpreted as something ($\{ \sigma _{\lambda _0 ...\lambda _{k}})\} $) which determines how local product bundles \lq glue together\rq . However, a product bundle is not just a line bundle, but a line bundle with a choice of section. To avoid introducing additional structure, we will therefore stick with pairs. 

Certainly, it is possible to drop the condition that $\{ U_{\lambda }\}$ should be a good covering as long as it is assumed that the line bundles $ \cL _{\lambda _0 ...\lambda _{k-1}}$ and $ (\delta \cL )_{\lambda _0 ...\lambda _{k}}$ are trivial. In this case we obtain a cohomology group $H^k(\{ U_\lambda \} ,\cL ,\sigma )$ which is isomorphic to $\check{H}^k(\{ U_\lambda \} ,\underline{{\bf C}}^\times )$. Let us point out that the constructions in this section goes through also if the line bundles are replaced by principal bundles with abelian fibres, i.e. $\underline{{\bf C}}^\times $ can be replaced with any other abelian group. Notice that $\delta \cL$ in general has a different fibre than $\cL$ in this case. 

\section{The \v{C}ech-de Rham complex} 
 The only assumptions that will be made in this, and the forthcoming sections, is that $B$ admits a partition of unity subordinate to a good covering if it is finite dimensional. If $B$ is infinite dimensional, some additional assumptions will be needed, see \cite{BE}. These assumptions will however not be discussed here. 
 
 It is well-known that if ignoring the torsion part in $\check{H}^k(B,\underline{{\bf C}}^\times )$, then it is isomorphic to a de Rham cohomology group. We will here review how the isomorphism is established by use of the  \v{C}ech-de Rham complex. This idea is very nice since it does not only prove the isomorphism, but it also shows that the \lq path\rq $ $ between the two cohomologies can be made step-wise. When \lq walking\rq $ $ through the  \v{C}ech-de Rham complex the degree of the \v{C}ech cocycles are decreased to the expense of an equal increase of the form degree (or vice versa). There will appear intermediate groups which are built up by elements which are mixtures of closed de Rham forms and \v{C}ech cocycles. This can be of advantage for us when trying to understand $\check{H}^k(B,\underline{{\bf C}}^\times )$. For instance, closed global 3-forms can be turned into a set of closed local 2-forms. This gives a geometrical realization of the free part of $\check{H}^2(B,\underline{{\bf C}}^\times )$ since closed local 2-forms can be considered as curvatures of local line bundles. 

 The abelian groups $\check{H}^k(B,\underline{{\bf C}})$ and $\check{H}^k(B,{\bf Z})$ are defined in the same way as $\check{H}^k(B,\underline{{\bf C}}^\times )$ but with ${\bf C}$- and ${\bf Z}$-valued cochains and the group operations given by addition.
 The reason for the notations $\underline{{\bf C}}$ and $\underline{{\bf C}}^\times $ is to differ from $\check{H}^k(B,{{\bf C}})$ and $\check{H}^k(B,{{\bf C}}^\times )$ which are defined in terms of
 cochains that are sets of constant functions on intersections. For ${\bf Z}$ this is unimportant since a smooth ${\bf Z}$-valued function over a contractible space is constant. Notice that since the sheaf of ${\bf C}$-valued functions is fine (there exist a partition of unity), it follows that $\check{H}^k(B,\underline{{\bf C}})=0$:
\eq
\label{eq:QE}
(\delta \omega )_{\lambda _0 ... \lambda _{k+1}}=0\Leftrightarrow \omega _{\lambda _0 ... \lambda _{k}}=(\delta\sum _\mu \rho_\mu \omega _{\mu _0...\mu _{k-1}})_{ \lambda _0 ... \lambda _{k}}.
\eqend

 Let $\{ g_{\lambda _0...\lambda _k} \}$ represent an element in $\check{H}^k(B,\underline{{\bf C}}^\times )$. Since $g_{\lambda _0...\lambda _k} $ is a ${\bf C}^\times$-valued function over
 a contractible set $U_{\lambda _0...\lambda _k} $, by choosing a branch of the logarithm, $\log g_{\lambda _0...\lambda _k} $ is a well-defined ${\bf C}$-valued function.
 However, although $\{ \log g_{\lambda _0...\lambda _k} \}$ is a \v{C}ech cochain, the logarithm may destroy the cocycle property. The obstruction is given by ${\bf Z}$-valued functions $\frac{1}{2\pi i} (\delta \log g)_{\lambda _0...\lambda _{k+1}}$ which themselves fulfill the cocycle condition and represents an element in $\check{H}^{k+1}(B,{\bf Z})$. This defines an isomorphism between $\check{H}^k(B,\underline{{\bf C}}^\times )$ and $\check{H}^{k+1}(B,{\bf Z})$. 

 It is well-known that there is a (non-canonical) splitting $\check{H}^{k+1}(B,{\bf Z})=\check{H}^{k+1}_0(B,{\bf Z})\oplus T^{k+1}_{\bf Z}(B)$, where we let $\check{H}^{k+1}_0(B,{\bf Z})={\bf Z}\oplus ...\oplus {\bf Z}$ denote the free part and $T^{k+1}_{\bf Z}(B)$ the torsion. The torsion vanishes when the coefficient field is extended from ${\bf Z}$ to ${\bf C}$, i.e. it is the kernel of the natural homomorphism from $\check{H}^{k+1}(B,{\bf Z})$ to $\check{H}^{k+1}(B,{\bf C})$.
 Clearly the kernel consist of the cocycles of $\check{H}^{k+1}(B,{\bf Z})$ which can be written as a coboundary with respect to constant ${\bf C}$-valued functions, i.e. a coboundary with respect to $\check{H}^{k+1}(B,{\bf C})$. Since every cocycle with respect to $\check{H}^{k+1}(B,{\bf Z})$ can be written as   $\frac{1}{2\pi i} (\delta \log g)_{\lambda _0...\lambda _{k+1}}$ we see that the kernel corresponds to constant $\log g_{\lambda _0...\lambda _{k}}$, or equivalently constant $g_{\lambda _0...\lambda _{k}}$.
 Thus, under the isomorphism between $\check{H}^k(B,\underline{{\bf C}}^\times )$ and $\check{H}^{k+1}(B,{{\bf Z}} )$, the torsion $T^{k+1}_{\bf Z}(B)$ corresponds to the abelian group $T^{k}_{\underline{{\bf C}}^\times  }(B)\subset \check{H}^k(B,\underline{{\bf C}}^\times )$ given by the constant cocycles $\{ g_{\lambda _0...\lambda _k}\} $.
 Notice that the cocycles in $T^{k}_{\underline{{\bf C}}^\times  }(B)$ and $\check{H}^k(B,{{\bf C}}^\times )$ agree but since the cochains differ, so does the coboundaries.

 We will now study the \v{C}ech-de Rham complex and the corresponding cohomology $\check{H}^{k+1}_{dR}(B,\underline{{\bf C}})$. A \v{C}ech-de Rham cochain of total degree $k$ is a set of ${\bf C}$-valued degree $q$ forms on every $U_{\lambda _0 ...\lambda _{p}}$ for $p+q=k+1$.
 The coboundary operator is $\delta +(-1)^pd$, where the domain of $\delta$ has been naturally extended to contain forms as well. The essential ideas concerning the \v{C}ech-de Rham complex are independent of $k$ and we will therefore take $k=2$ as an illustrating example.
 Let $F\in\Omega ^3(B,\underline{{\bf C}})$ be a closed 3-form on $B$. Then $\delta$ maps $F$ into the \v{C}ech-de Rham complex according to $\delta F=\{ F_\lambda \}$, where $F_\lambda =F|_{U_\lambda }$. Since $dF=0$ we can use the Poincar\'{e} lemma on $U_\lambda $ to write $(\delta F )_\lambda =F_\lambda =d\ommega _\lambda$.
 Then the Poincar\'{e} lemma can be used once more on $0=(\delta ^2F)_{\lambda\mu }=(\delta d\ommega  )_{\lambda\mu }=(d\delta \ommega  )_{\lambda\mu }$ to give $(\delta \ommega  )_{\lambda\mu }=d\ommmega _{\lambda\mu }$. The Poincar\'{e} lemma on $0=(\delta ^2\ommega  )_{\lambda\mu\nu }=(\delta d\ommmega )_{\lambda\mu\nu }=(d\delta \ommmega )_{\lambda\mu\nu }$ gives $(\delta \ommmega )_{\lambda\mu\nu }=dc_{\lambda\mu\nu }$.
 Since $0=(\delta ^2\ommmega )_{\lambda\mu\nu\epsilon }=(\delta dc)_{\lambda\mu\nu\epsilon  }=(d\delta c)_{\lambda\mu\nu\epsilon  }$, the functions $(\delta c)_{\lambda\mu\nu\epsilon  }$ must be constant and thereby defining a cocycle $c_{\lambda\mu\nu\epsilon  }$
 that represents an element in $\check{H}^3(B,{{\bf C}})$. It is easy \cite{BT} to see that this procedure gives isomorphisms $H^{k+1}_{dR}(B,\underline{{\bf C}})\cong\check{H}^{k+1}_{dR}(B,\underline{{\bf C}})\cong\check{H}^{k+1}(B,{{\bf C}})$ for $k=2$ and by
 the generality of the method this is true for all $k$. The abelian group structure in $\check{H}^{k+1}_{dR}(B,\underline{{\bf C}})$ is defined in the obvious way.
\[
\begin{array}{ccccccccc}
 \Omega ^3 && F_\lambda & \longrightarrow & 0 &&&&\nonu
           &&  \uparrow &                 & \uparrow  &&&&\nonu
 \Omega ^2 && \ommega _\lambda & \longrightarrow & (\delta \ommega  )_{\lambda\mu } & \longrightarrow & 0 &&\nonu
&&&&      \uparrow  && \uparrow     &&\nonu
 \Omega ^1 &&&& \ommmega _{\lambda\mu } & \longrightarrow & (\delta \ommmega )_{\lambda\mu\nu } & \longrightarrow & 0 \nonu
&&&&&&  \uparrow   &&  \uparrow  \nonu
\Omega ^0&&&&&& c_{\lambda\mu\nu } & \longrightarrow &  (\delta c)_{\lambda\mu\nu\epsilon }            \nonu
&&&&&&&&\nonu
d\uparrow \stackrel{\delta }{\longrightarrow  }&&C^0&&C^1&&C^2&&C^3\nonu
\end{array}
\]
 The main results so far in this section can be summarized in the above and in the following diagram:
\eq
\begin{array}{ccccccc}
\label{di:ISO}
0  \longrightarrow & T^{k}_{\underline{{\bf C}}^\times }(B) & \longrightarrow & \check{H}^{k}(B,\underline{{\bf C}}^\times ) & \longrightarrow & \check{H}^{k}_0(B,\underline{{\bf C}}^\times )   &   \longrightarrow \quad 0\nonu
&\cong &&\cong && \cong & \nonu
0  \longrightarrow & T^{k+1}_{\bf Z}(B) & \longrightarrow & \check{H}^{k+1}
(B,{\bf Z}) & \longrightarrow & \check{H}^{k+1}_0(B,{\bf Z} ) & \longrightarrow \quad 0 \nonu
& && && \cong &\nonu
 & & &  && \check{H}^{k+1}_0(B,{\bf C})  & \subset  \check{H}^{k+1}(B,{\bf C}) \nonu
&&&&& \cong & \cong \nonu
&&&&& \check{H}^{k+1}_{dR,0}(B,\underline{{\bf C}})  & \subset  \check{H}^{k+1}_{dR}(B,\underline{{\bf C}}) \nonu
&&&&& \cong & \cong \nonu
&&&&& H^{k+1}_{dR,0}(B,\underline{{\bf C}})  & \hspace{1mm}\subset  H^{k+1}_{dR}(B,\underline{{\bf C}}).
\end{array}
\eqend
 Here $\check{H}^{k+1}_{0}(B,{\bf C}) $ denotes the image of $\check{H}^{k+1}(B,{\bf Z})$ in $\check{H}^{k+1}(B,{\bf C}) $ and the two last groups in the second column from the right are induced from the ones in the right column.

 Consider now the map from $\check{H}^{k}(B,\underline{{\bf C}}^\times )$ to $H^{k+1}_{dR}(B,\underline{{\bf C}})$ given by the diagram. We then have to \lq walk\rq $ $ through the \v{C}ech-de Rham complex from the bottom right to the upper left.
 This is possible since there is a \lq Poincar\'{e} lemma\rq $ $ for $\delta $ as well, see eq. \Ref{eq:QE}. The equations that appears when walking through the \v{C}ech-de Rham complex takes the following form for $k=1$:
\[
\left\{ \begin{array}{ccccc}
F_\lambda & = & d\omega _\lambda && \nonu
\omega _\lambda - \omega _\mu & = & dc_{\lambda \mu } & = & d\log g_{\lambda \mu }.
\end{array}\right.
\]
 We thus see that if the $g_{\lambda \mu }$'s are interpreted as transition functions of a line bundle, then the $\omega _\lambda$'s can be interpreted as connection 1-forms on the base manifold and $F$ as the curvature.
 Therefore,  $\check{H}^{1}(B,\underline{{\bf C}}^\times )\rightarrow H^{2}_{dR}(B,\underline{{\bf C}})$ is the map from the class of a line bundle to the cohomology class (the Chern class) of its curvature. The kernel $T^{1}_{\underline{{\bf C}}^\times }(B)$ of this homomorphism is important.
 Its existence implies that two non-equivalent line bundles can have equal characteristic classes. In fact, $T^{1}_{\underline{{\bf C}}^\times }(B)$ is precisely the holonomy group of the line bundle. It is well-known that $[F]\in {H}^{1}_{dR,0}(B,\underline{{\bf C}})$ is equivalent with that $F$ fulfills the integrality condition. In the Appendix we will prove the generalization of this statement:
\begin{proposition}
\label{p:2}
$[F]\in {H}^{k+1}_{dR,0}(B,\underline{{\bf C}})\Leftrightarrow$ $F$ satisfies the integrality condition.
\end{proposition}
For $k=2$ there exist a similar terminology as for $k=1$: one speaks about connective structures, curvings and holonomy, \cite{BR}.

 In section 2 we obtained a realization of the elements in $\check{H}^{k}(B,\underline{{\bf C}}^\times )$ in terms of sets of pairs of line bundles and sections. Since $\check{H}^{k+1}_{dR,0}(B,\underline{{\bf C}})$ is closely related according to the diagram in \Ref{di:ISO}, there might exist a similar realization for this group. We will show that this is the case for $\check{H}^{k+1}_{dR,0}(B,\underline{{\bf C}})_{(k,1)}$ and $\check{H}^{k+1}_{dR,0}(B,\underline{{\bf C}})_{(k-1,2)}$ (which both are isomorphic to $\check{H}^{k+1}_{dR,0}(B,\underline{{\bf C}})$). The index $(p,q)$ refers to the part of the \v{C}ech-de Rham cohomology that is built up by q-forms on each $U_{\lambda _0...\lambda _{p}}$. Recall that $\check{H}^{k+1}_{dR}(B,\underline{{\bf C}})_{(p,k-p+1)}\cong\check{H}^{k+1}_{dR}(B,\underline{{\bf C}})$ We will consider sets $\{ (\cL _{\lambda _0...\lambda _{k-1}},\nabla _{\lambda _0...\lambda _{k-1}}) \}$ of pairs of line bundles with connections. As usual, the line bundles $\cL _{\lambda _0...\lambda _{k-1}}$ are equipped with isomorphisms $\xi _{\lambda _0...\lambda _{k-1}}$ to their duals.
\begin{definition}
\label{d:4}
$\{ (\cL _{\lambda _0 ...\lambda _{k-1}}, \nabla _{\lambda _0 ...\lambda _{k-1}})\}$ and $\{ (\cL ^\prime _{\lambda _0 ...\lambda _{k-1}}, \nabla ^\prime _{\lambda _0 ...\lambda _{k-1}})\}$ are said to be equivalent if there exist a diffeomorphism between them, i.e. diffeomorphisms $\varphi _{\lambda _0 ...\lambda _{k-1}}: \cL _{\lambda _0 ...\lambda _{k-1}}\rightarrow \cL ^\prime _{\lambda _0 ...\lambda _{k-1}}$ such that 
\[
\varphi _{\lambda _0 ...\lambda _{k-1}}(\nabla _{\lambda _0 ...\lambda _{k-1},v}s_{\lambda _0 ...\lambda _{k-1}})=\nabla ^\prime _{\lambda _0 ...\lambda _{k-1},v}\varphi _{\lambda _0 ...\lambda _{k-1}}(s_{\lambda _0 ...\lambda _{k-1}}),
\]
for any tangent field $v$ on $B$ and any section $s_{\lambda _0 ...\lambda _{k-1}}$ on $\cL _{\lambda _0 ...\lambda _{k-1}}$.
\end{definition}
 Multiplication of pairs gives by definition $\{ (\cL _{\lambda _0 ...\lambda _{k-1}}\otimes \cL ^\prime _{\lambda _0 ...\lambda _{k-1}}, \nabla _{\lambda _0 ...\lambda _{k-1}}+\nabla ^\prime _{\lambda _0 ...\lambda _{k-1}})\}$, where 
\begin{eqnarray*}
&&\nabla _{\lambda _0 ...\lambda _{k-1}}+\nabla ^\prime _{\lambda _0 ...\lambda _{k-1}}(s_{\lambda _0 ...\lambda _{k-1}}\otimes s ^\prime _{\lambda _0 ...\lambda _{k-1}} )\nonu
&&= (\nabla _{\lambda _0 ...\lambda _{k-1}}s_{\lambda _0 ...\lambda _{k-1}})\otimes s ^\prime _{\lambda _0 ...\lambda _{k-1}}+s_{\lambda _0 ...\lambda _{k-1}}\otimes (\nabla ^\prime _{\lambda _0 ...\lambda _{k-1}}s ^\prime _{\lambda _0 ...\lambda _{k-1}} ). \nonu
\end{eqnarray*}
 This defines an abelian group $C^{k+1}(\{ U_\lambda \} ,\cL , \nabla )$ whose elements are the equivalence classes of (sets of) pairs. The identity element clearly consist of sets of pairs $\{ (\cL _{\lambda _0 ...\lambda _{k-1}}, \nabla _{\lambda _0 ...\lambda _{k-1}})\}$ such that there exist non-vanishing horizontal (flat) sections, i.e. sections satisfying $\nabla _{\lambda _0 ...\lambda _{k-1}}s_{\lambda _0 ...\lambda _{k-1}}=0$. Every element in $C^{k+1}(\{ U_\lambda \} ,\cL , \nabla )$ 
defines a \v{C}ech-de Rham cochain $\{ (\delta \omega )_{\lambda _0 ...\lambda _{k}}\}$ up to a coboundary by
\eq
\label{eq:BI2}
\nabla _{\lambda _0 ...\lambda _{k-1}}s_{\lambda _0 ...\lambda _{k-1}}=s_{\lambda _0 ...\lambda _{k-1}}\omega _{\lambda _0 ...\lambda _{k-1}},
\eqend
where $s_{\lambda _0 ...\lambda _{k-1}}$ is a non-vanishing section of $\cL _{\lambda _0 ...\lambda _{k-1}}$.
 The map 
\[
(\cL _{\lambda _0 ...\lambda _{k-1}},\nabla _{\lambda _0 ...\lambda _{k-1}})\mapsto ((\delta \cL )_{\lambda _0 ...\lambda _{k}},(\delta \nabla  )_{\lambda _0 ...\lambda _{k}})\nonumber
\]
 induces a homomorphism $\delta : C^{k+1}(\{ U_\lambda \} ,\cL ,\nabla  )\rightarrow C^{k+2}(\{ U_\lambda \} ,\cL ,\nabla  )$. The operator $(\delta \nabla  )_{\lambda _0 ...\lambda _{k}}$ is defined in the obvious way. Notice that $(\delta ^{2}s )_{\lambda _0 ...\lambda _{k+1}}$ is horizontal with respect to $(\delta ^{2}\nabla  )_{\lambda _0 ...\lambda _{k+1}}$. Since $\delta ^2$ gives the identity element we are lead to:
\begin{definition}
\label{d:41}
 Let $H^{k+1}(\cL ,\nabla  )$ be the cohomology defined by $C^{k+1}(\{ U_\lambda \} ,\cL ,\nabla  )$ and $\delta $. $H^{k+1}_0(\cL ,\nabla  )$ denotes the subgroup built up by pairs  with horizontal sections $\sigma _{\lambda _0 ...\lambda _{k}}$ of $(\delta \cL )_{\lambda _0 ...\lambda _{k}}$ with respect to $(\delta \nabla )_{\lambda _0 ...\lambda _{k}}$ such that $(\delta \sigma )_{\lambda _0 ...\lambda _{k+1}}$ equals the canonical section $(\delta ^2s )_{\lambda _0 ...\lambda _{k+1}}$.
\end{definition}
 Eq. \Ref{eq:BI2} gives $H^{k+1}(\cL ,\nabla )\cong\check{H}^{k+1}_{dR}(B,\underline{{\bf C}})_{(k,1)}$ and restricts to $H^{k+1}_0(\cL ,\nabla )\cong\check{H}^{k+1}_{dR,0}(B,\underline{{\bf C}})_{(k,1)}$. The latter isomorphism can be proven by using the fact that $H^{k+1}_0(\cL ,\nabla )$ is isomorphic to the free part $H^k_0(\cL ,\sigma )$ of $H^k(\cL ,\sigma )$. To show this it is helpful to use that $H^k_0(\cL ,\sigma )\cong\check{H}^{k}_0(B,{\underline{{\bf C}}^\times })$, where the latter group is $\check{H}^{k}(B,{\underline{{\bf C}}^\times })$ modulo constant cocycles. Every $[\{ (\cL _{\lambda _0 ...\lambda _{k-1}},\nabla _{\lambda _0 ...\lambda _{k-1}})\}]\in H^{k+1}(\cL ,\nabla )$ defines an element $[\{ F_{\lambda _0 ...\lambda _{k-1}}\} ]\in \check{H}^{k+1}_{dR}(B,\underline{{\bf C}})_{(k-1,2)}$ by:
\eq
(\nabla ^2)_{\lambda _0 ...\lambda _{k-1}}s_{\lambda _0 ...\lambda _{k-1}}=s_{\lambda _0 ...\lambda _{k-1}}F_{\lambda _0 ...\lambda _{k-1}},
\eqend
where $(\nabla ^2)_{\lambda _0 ...\lambda _{k-1}}$ is the curvature of $\nabla _{\lambda _0 ...\lambda _{k-1}}$. This induces an isomorphism $H^{k+1}(\cL ,\nabla )\cong\check{H}^{k+1}_{dR}(B,\underline{{\bf C}})_{(k-1,2)}$, which in turn restricts to $H^{k+1}_0(\cL ,\nabla )\cong\check{H}^{k+1}_{dR,0}(B,\underline{{\bf C}})_{(k-1,2)}$. 
We summarize the isomorphisms above in the following diagram:
\begin{lemma}
\label{l:1}
\[
\begin{array}{ccccc}
H^k_0(\cL ,\sigma ) & \cong & H^{k+1}_0(\cL ,\nabla ) & = & H^{k+1}_0(\cL ,\nabla ) \nonu
\cong && \cong &&\cong \nonu
\check{H}^{k}_0(B,{\underline{{\bf C}}^\times }) & \cong & \check{H}^{k+1}_{dR,0}(B,\underline{{\bf C}})_{(k,1)} & \cong & \check{H}^{k+1}_{dR,0}(B,\underline{{\bf C}})_{(k-1,2)} .
\end{array}
\]
\end{lemma}

\section{The pull-back to an affine space}
 Let $\A$ be an affine space and $\pi :\A\rightarrow B$ a fibre bundle. We will see how some parts of the \v{C}ech-de Rham complex can be pulled-back from $B$ to $\A$. The reason why not all elements in the complex can be pulled-back is that $\{ \pi ^{-1}(U_\lambda )\} _{\lambda \in \Lambda }$ is not necessary a good covering. The advantage of the pull-back is that the form degrees will be of one order less, see below. Thus, the case $k=2$ can then almost be treated as the simple case when $k=1$. Some difficulties will remain however. Unfortunately, the case $k=3$ will not become simpler than what the case $k=2$ was from the beginning. Focus will therefore be on the cases $k=1$ and $k=2$. An interesting feature is that the integrality condition is not invariant under such a pull-back (or transgression). This comes from the existence of lower homotopy groups in the base manifold $B$.
 Often, we will put a hat on symbols referring to line bundles, forms, etc., when they are defined with respect to $\A$. Exceptions of this rule are the forms $\hat{a}$ and $\hat{\eta }$ that appears in the next section.
\begin{definition}
\label{d:5}
 Denote by $\Omega ^{k}(\A \stackrel{\pi }{\rightarrow }B,\underline{{\bf C}})$ the quotient group of $\Omega ^k(\A ,\underline{{\bf C}})$ by forms of type $\pi ^\ast\omega $. Then $H^{k}_{dR}(\A \stackrel{\pi }{\rightarrow }B,\underline{{\bf C}})$ is the cohomology defined by $\Omega ^{k}(\A \stackrel{\pi }{\rightarrow }B,\underline{{\bf C}})$ and the coboundary operator (induced by) $\hat{d}$. Let $H^{k}_{dR,0}(\A \stackrel{\pi }{\rightarrow }B,\underline{{\bf C}})$ be the subgroup of representatives such that the integral over any closed $k$-manifold in a fibre gives an element in $2\pi i{\bf Z}$. 
\end{definition}
\begin{lemma}
\label{l:2}
$H^{k}_{dR}(\A \stackrel{\pi }{\rightarrow }B,\underline{{\bf C}})\cong H^{k+1}_{dR}(B,\underline{{\bf C}})$
\end{lemma}
\vspace{2mm}
{\bf Proof}\hspace{2mm}
 For $[F]\in H^{k+1}_{dR}(B,\underline{{\bf C}})$ the Poincar\'{e} lemma gives the existence of an $[\hat{\omega }]\in H^{k}_{dR}(\A \stackrel{\pi }{\rightarrow }B,\underline{{\bf C}})$ so that $\hat{d} \hat{\omega }=\pi ^\ast F$. The ambiguity in the choice of $\hat{\omega }$ and triviality of $F$ both give coboundaries so a map $H^{k+1}_{dR}(B,\underline{{\bf C}})\rightarrow H^{k}_{dR}(\A \stackrel{\pi }{\rightarrow }B,\underline{{\bf C}})$ has been obtained. Since it is linear, surjective and invertible it is also an isomorphism. 
\eop
\begin{lemma}
\label{l:25}
There exist a monomorphism $H^{k+1}_{dR,0}(B,\underline{{\bf C}})\rightarrow H^{k}_{dR,0}(\A \stackrel{\pi }{\rightarrow }B,\underline{{\bf C}})$
\end{lemma}
\vspace{2mm}
{\bf Proof}\hspace{2mm}
The monomorphism is the isomorphism in lemma \ref{l:2}, but with a different domain and range. The only thing that needs to be checked is that $\hat{\omega }$ satisfies the integrality condition if $F$ does it. Let therefore $S$ be an arbitrary closed $k$-manifold contained in a fibre. Since $\A $ is affine there exist a manifold $V$ with $\partial V =S$. Then:
\[
\int _S\hat{\omega } =\int _V\pi ^\ast F=\int _{\pi (V)} F
\]
which is in $2\pi i{\bf Z}$ since $\pi (V)$ is a manifold without boundary.
\eop

The proof above uses ideas from a proof of the exact homotopy sequence of a fibre bundle, see \cite{SW}, chapter 4. For instance, the following fact is important: The group of homotopy classes of maps from the k-sphere: $S^k\rightarrow B$ is in bijective correspondence with the group of homotopy classes of maps $D^{k+1}\rightarrow \A$ such that $\partial D^{k+1}$ is mapped into a fibre, see figure below:

\vspace{0.05cm}

\begin{center}\epsfxsize=1.5cm
$$\hspace{-3cm}\hbox{\epsfbox{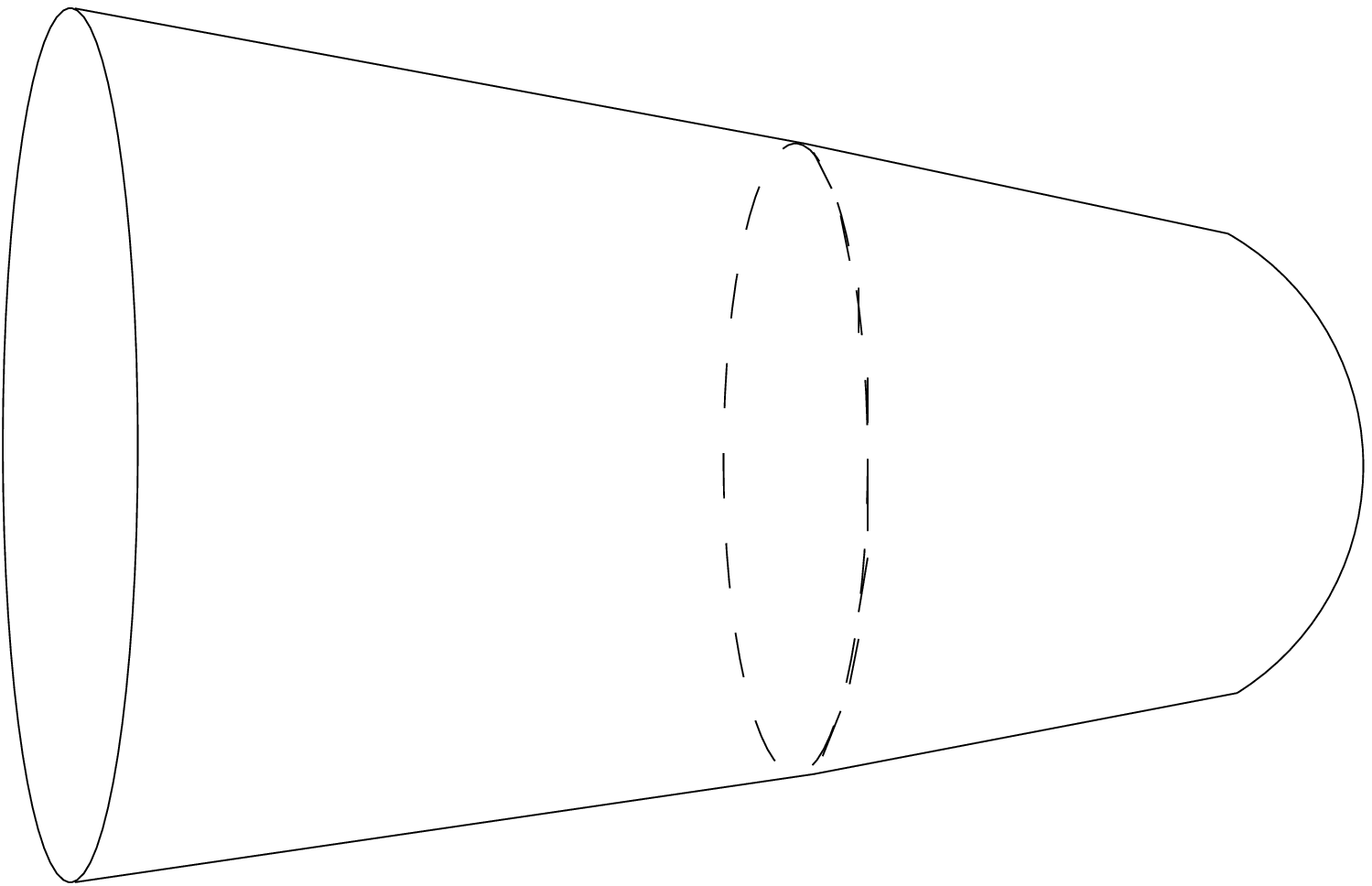}}$$
\vspace{0.5cm}
\epsfxsize=1.5cm
$$\hspace{-3cm}\hbox{\epsfbox{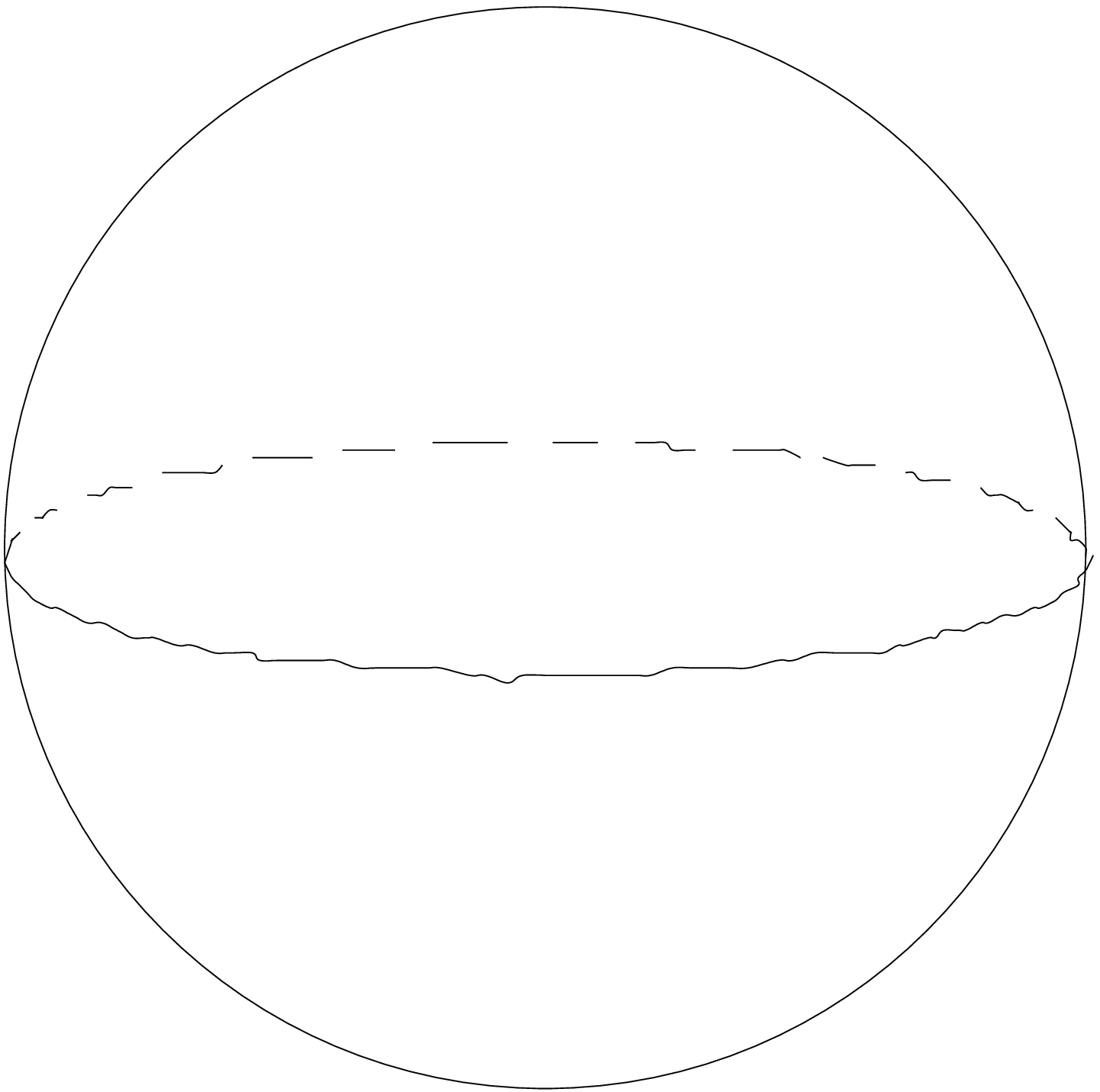}}$$
\end{center}
\begin{picture}(0,0)(0,0)
\put(190,115){$\subset$}
\put(220,115){$\A$}
\put(125,80){$\downarrow$}
\put(220,80){$\downarrow$}
\put(99,150){\footnotesize{fibre}}
\put(105,138){$\cup$}
\put(190,45){$\subset$}
\put(220,45){$B$}
\end{picture}

The integral of $F$ over a manifold $\pi (V)$ can only be non-zero if there doesn't exist any manifold which has $\pi (V)$ as its boundary (since $F$ is closed). The reason why we only obtained a monomorphism in lemma \ref{l:25} is that there are \lq more\rq $ $ $(k+1)$-manifolds on $B$ of this type than $k$-manifolds in the fibre of the same type. This happens even for $k=1$: Every 2-sphere on $B$ lifts to a bowl on $\A $ which restricts to a 1-sphere in the fibre, and vice versa. This proves that $\pi _1(\mbox{fibre})=\pi _2(B)$. However, there are more types of manifolds on $B$ that are obstructions in that the integral of $F$ should be zero. This comes from the existence of lower homotopy groups. Consider for example the manifold $B$ pictured below. It looks like an apple where a worm has made a hole from the top to the bottom and then (somehow) a torus around the original hole. By taking $\pi (V)$ as the surface of the torus we see that there is no closed manifold in the fibre that corresponds to it. To summarize: there is an injective map from the set of homotopy classes of closed $k$-manifolds in a fibre to the set of homotopy classes of closed $(k+1)$-manifolds on the base. 

\begin{center}\epsfxsize=2.5cm
$$\hbox{\epsfbox{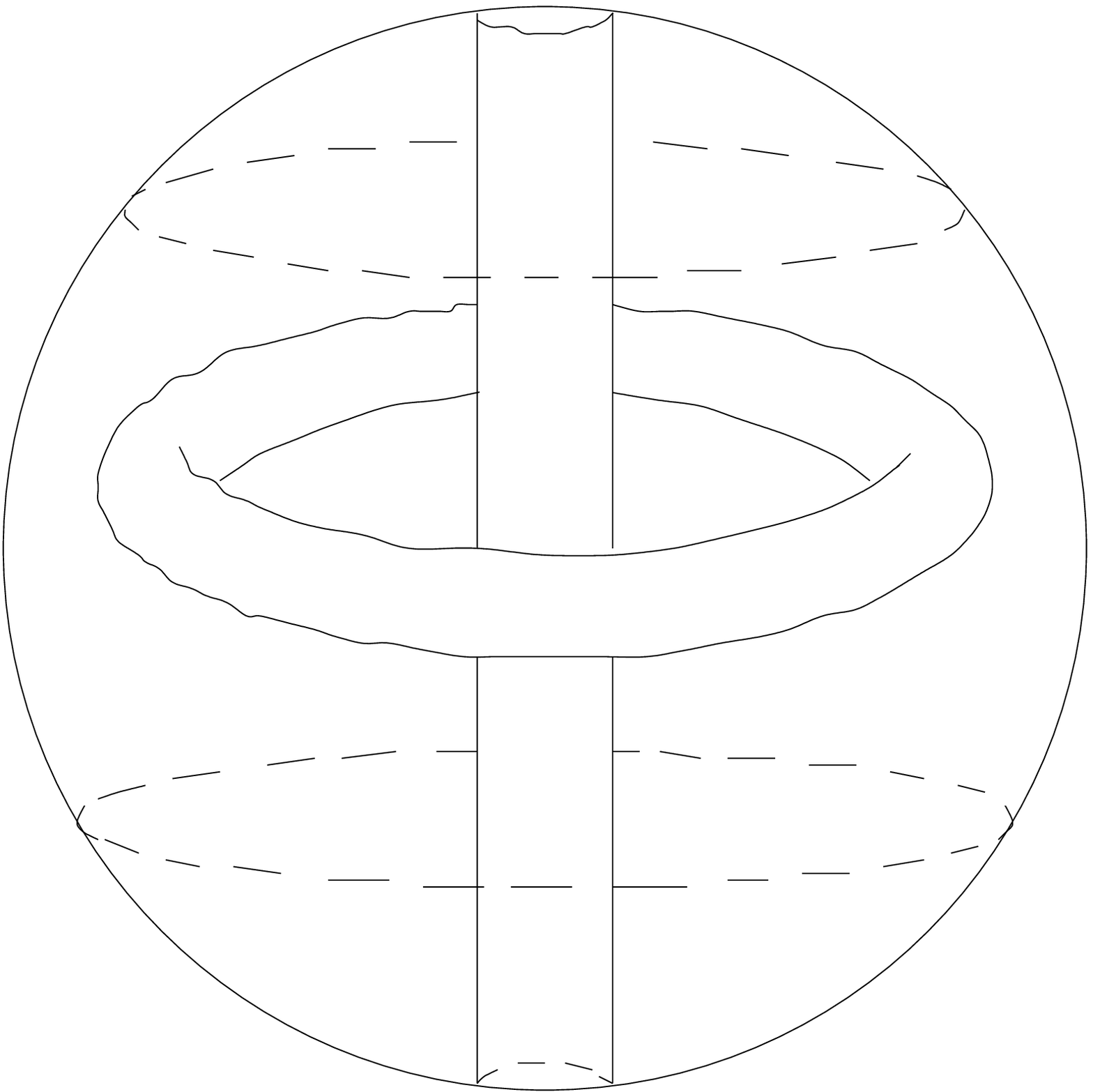}}$$
\end{center}

The discussion above is not completely honest. When considering homotopy groups  one also have to take into account the dependence of the (base-) points for which the homotopy group is defined, \cite{SW}. However, when the space is connected there is no such dependence. Clearly, $B$ is path-connected: take two points in $B$, lift them to the affine space $\A$ and then project down a path joining the two lifted points. For the fibre there is no corresponding trick, so here we will get a dependence on the choice of base point for the homotopy groups.
\begin{definition}
\label{d:6}
 Denote by $\tilde{\Omega }^{k}(\{ U_\lambda\} , \A \stackrel{\pi }{\rightarrow }B,\underline{{\bf C}})$ the abelian group of elements $\{ \tilde{\omega }_\lambda \}$ consisting of $k$-forms over $\pi ^{-1}(U_\lambda )$ of type:
\eq
\label{eq:11}
\tilde{\omega }_\lambda =(\hat{\delta }\hat{\omega })_\lambda -\pi ^\ast \omega _\lambda .
\eqend
 The cohomology $\tilde{H }^{k}(\A \stackrel{\pi }{\rightarrow }B,\underline{{\bf C}})$ is then defined by $\tilde{\Omega }^{k}(\{ U_\lambda\} , \A \stackrel{\pi }{\rightarrow }B,\underline{{\bf C}})$ and the coboundary operator $\hat{d}$. Define $\tilde{H }^{k}_0(\A \stackrel{\pi }{\rightarrow }B,\underline{{\bf C}})$ to be the subgroup of representatives $\tilde{\omega }_\lambda $ that satisfies the integrality condition with respect to $k$-manifolds in $\pi ^{-1}(U_\lambda )$.
\end{definition}
\begin{lemma}
\label{l:3}
$\tilde{H}^{k}(\A \stackrel{\pi }{\rightarrow }B,\underline{{\bf C}})$ is isomorphic to each of the grops $\check{H}^{k+1}_{dR}(B,\underline{{\bf C}})_{(1,k)}$,\linebreak $\check{H}^{k+1}_{dR}(B,\underline{{\bf C}})_{(0,k+1)}$ and ${H}^{k}_{dR}(\A \stackrel{\pi }{\rightarrow }B,\underline{{\bf C}})$.
\end{lemma}
\vspace{2mm}
{\bf Proof}\hspace{2mm}
Applying $\hat{d} $ on a representative $\{ \tilde{\omega }_\lambda \}$ of $\tilde{H}^{k}(\A \stackrel{\pi }{\rightarrow }B,\underline{{\bf C}})$ gives: 
\[
(\hat{\delta }\hat{d} \hat{\omega })_\lambda =\pi ^\ast d\omega _\lambda .
\]
 It implies that $\pi ^\ast (\delta d\omega )_{\lambda\mu }=0$ so $d\omega _\lambda =(\delta F )_\lambda $ for some $F$. Thus  $[\{ d\omega _\lambda \} ]\in \check{H}^{k+1}_{dR}(B,\underline{{\bf C}})_{(0,k+1)}$ and  $[\{ (\delta\omega )_{\lambda\mu } \} ]\in \check{H}^{k+1}_{dR}(B,\underline{{\bf C}})_{(1,k)}$. The equation above then becomes $(\hat{\delta }\hat{d} \hat{\omega })_\lambda =(\hat{\delta }\pi ^\ast F )_\lambda $, or equivalently $\hat{d} \hat{\omega }=\pi ^\ast F $, which implies that $[\hat{\omega }]\in {H}^{k}_{dR}(\A \stackrel{\pi }{\rightarrow }B,\underline{{\bf C}})$.  It is straight forward to check that this gives the claimed isomorphisms. We will be settled by showing that the isomorphisms are independent of the decomposition in \Ref{eq:11}. Let $\tilde{\omega }_\lambda =(\hat{\delta }\hat{\omega }^\prime )_\lambda -\pi ^\ast \omega _\lambda ^\prime $ be a different decomposition and set $\triangle \hat{\omega }=\hat{\omega }^\prime -\hat{\omega }$ and $\triangle \omega _\lambda = \omega _\lambda ^\prime -\omega _\lambda $. Then 
\[
\pi ^\ast \sum _\lambda \rho _\lambda \triangle \omega _\lambda = \sum _\lambda \pi ^\ast \rho _\lambda \pi ^\ast \triangle \omega _\lambda = \sum _\lambda \pi ^\ast \rho _\lambda (\hat{\delta }\triangle \hat{\omega })_\lambda 
= \triangle \hat{\omega }  \sum _\lambda \pi ^\ast \rho _\lambda =\triangle \hat{\omega }  
\]
which shows that $\hat{\omega }^\prime$ and $\hat{\omega }$ represents the same element in ${H}^{k}_{dR}(\A \stackrel{\pi }{\rightarrow }B,\underline{{\bf C}})$. Further, 
\[
\pi ^\ast \triangle \omega _\lambda =(\hat{\delta }\triangle \hat{\omega })_\lambda =\pi ^\ast (\hat{\delta }\sum _\mu \rho _\mu\triangle \omega _\mu )_\lambda 
\] 
which shows that the images in $\check{H}^{k+1}_{dR}(B,\underline{{\bf C}})_{(1,k)}$ respective $\check{H}^{k+1}_{dR}(B,\underline{{\bf C}})_{(0,k+1)}$ are independent of the decomposition.
\eop

 Although $\{\pi ^{-1} (U_\lambda )\}$ is not a good covering we obtain as a corollary that $\tilde{H}^{k}(\A \stackrel{\pi }{\rightarrow }B,\underline{{\bf C}})$ is independent of the covering (as long as it is good on $B$).
\begin{lemma}
\label{l:35}
$\tilde{H}^{k}_0(\A \stackrel{\pi }{\rightarrow }B,\underline{{\bf C}})\cong {H}^{k}_{dR,0}(\A \stackrel{\pi }{\rightarrow }B,\underline{{\bf C}})$.
\end{lemma}
\vspace{2mm}
{\bf Proof}\hspace{2mm}
Let $S\subset\pi ^{-1}(U_\lambda )$ be a closed $k$-manifold. Then $\pi (S)$ is a closed manifold in the contractible set $U_\lambda $ and therefore there exist a manifold $\pi (V)$ with $\partial \pi (V)=\pi (S)$. Since $\pi (V)$ can be obtained from a manifold $V$ on $\pi ^{-1}(U_\lambda )$, the notation is justified. $V$ can be chosen such that $\partial V=S\cup S^\prime $, where $S^\prime$ is such that $\pi (S^\prime )$ is a point. With notations from the proof of lemma \ref{l:3} we get: 
\begin{eqnarray*}
\int _S\tilde{\omega }_\lambda & = & \int _S\left( (\hat{\delta }\hat{\omega })_\lambda -\pi ^\ast \omega _\lambda \right) =\int _V\pi ^\ast F -\int _{S^\prime }\hat{\omega }-\int _{\pi (S)}\omega _\lambda \nonu
& = & \int _{\pi (V)}F -\int _{S^\prime }\hat{\omega }-\int _{\pi (V)}F = -\int _{S^\prime }\hat{\omega }.
\end{eqnarray*}
The converse part of the proof is obvious.
\eop

 Putting together the \v{C}ech-de Rham complex with lemma \ref{l:1}--\ref{l:35}, the statement below follows:
\begin{theorem}
\label{t:1}
\[
\begin{array}{ccccccc}
& H^k_0(\cL ,\sigma ) & \cong & H^{k+1}_0(\cL ,\nabla ) & = & H^{k+1}_0(\cL ,\nabla ) &\nonu
& \cong && \cong &&\cong &\nonu
& \check{H}^{k}_0(B,{\underline{{\bf C}}^\times }) & \cong & \check{H}^{k+1}_{dR,0}(B,\underline{{\bf C}})_{(k,1)} & \cong & \check{H}^{k+1}_{dR,0}(B,\underline{{\bf C}})_{(k-1,2)} & \cong \nonu
\end{array}
\]
\[
\begin{array}{ccccccc}
\cong & \check{H}^{k+1}_{dR,0}(B,\underline{{\bf C}})_{(k-2,3)}& \cong & ... &\cong &\check{H}^{k+1}_{dR,0}(B,\underline{{\bf C}})_{(2,k-1)} & \cong \nonu
\end{array}
\]
\[
\begin{array}{ccccccc}
& \tilde{H}^{k}_0(\A \stackrel{\pi }{\rightarrow }B,\underline{{\bf C}})& = & \tilde{H}^{k}_0(\A \stackrel{\pi }{\rightarrow }B,\underline{{\bf C}})& \cong & {H}^{k}_{dR,0} (\A \stackrel{\pi }{\rightarrow }B,\underline{{\bf C}}) &\nonu
& \uparrow \mbox{\footnotesize{monic}}  && \uparrow \mbox{\footnotesize{monic}}  && \uparrow \mbox{\footnotesize{monic}}  & \nonu
\cong &\check{H}^{k+1}_{dR,0}(B,\underline{{\bf C}})_{(1,k)} &\cong & \check{H}^{k+1}_{dR,0}(B,\underline{{\bf C}})_{(0,k+1)} & \cong & {H}^{k+1}_{dR,0}(B,\underline{{\bf C}}). & 
\end{array}
\]
\end{theorem}
This becomes particular interesting when $k=1$ or 2. The reason is that there exist rather simple geometrical realizations of $H^{1}_{dR,0}(\A \stackrel{\pi }{\rightarrow }B,\underline{{\bf C}})$ and $\tilde{H }^{2}_0(\A \stackrel{\pi }{\rightarrow }B,\underline{{\bf C}})$. In the following definition we use that ${C}^{k}(\{ \pi ^{-1}(U_\lambda )\} ,\hat{\cL },\hat{\nabla })$, $k\geq 1$, can be defined also when the line bundles $\cL _{\lambda _0...\lambda _{k-2}}$ are non-trivial.
\begin{definition}
\label{d:7}
 Let $\tilde{C}^{k}(\{ \pi ^{-1}(U_\lambda )\} ,\hat{\cL },\hat{\nabla })$ be the quotient of ${C}^{k}(\{ \pi ^{-1}(U_\lambda )\} ,\hat{\cL },\hat{\nabla })$ by the group generated by elements which have a representative consisting of pairs of type $\{ (\pi ^{\ast}\cL _{\lambda _0 ... \lambda _{k-2}} ,\pi ^{\ast}\nabla _{\lambda _0 ... \lambda _{k-2}} )\}$. The cohomology $\tilde{H}^{k} (\hat{\cL },\hat{\nabla })$ is then defined by $\tilde{C}^{k}(\{ \pi ^{-1}(U_\lambda )\} ,\hat{\cL },\hat{\nabla })$ and the coboundary operator $\hat{\delta }$ induced by the corresponding operator acting on ${C}^{k}(\{ \pi ^{-1}(U_\lambda )\} ,\hat{\cL },\hat{\nabla })$. 
\end{definition}
 Let us first focus on the case $k=1$.
\begin{lemma}
\label{l:4}
$\tilde{H}^{1}(\hat{\cL },\hat{\nabla })\cong {H}^{1}_{dR}(\A \stackrel{\pi }{\rightarrow }B,\underline{{\bf C}})$
\end{lemma} 
\vspace{2mm}
{\bf Proof}\hspace{2mm}
 When $k=1$, the representatives of $\tilde{C}^{k}(\{ \pi ^{-1}(U_\lambda )\} ,\hat{\cL },\hat{\nabla })$ are line bundles $\hat{\cL }$ with connections $\hat{\nabla }$, globally defined on $\A $. It is trivial if $(\hat{\cL },\hat{\nabla })$ is equivalent with $(\pi ^\ast {\cL }, \pi ^\ast {\nabla })$ for some trivial bundle $\cL$ on $B$ with connection $\nabla $. Since $\A $ is affine, there exist a non-vanishing section $\hat{s}$ of $\hat{\cL }$. Define $\hat{\omega }$ by:
\eq
\label{eq:EQ}
\hat{\nabla }\hat{s}=\hat{s}\hat{\omega }.
\eqend
The cocycle condition implies that $\hat{d} \hat{\omega }=\pi ^\ast F$ for some $[F]\in {H}^{2}_{dR}(B,\underline{{\bf C}})$. It is straight forward to check that this defines an isomorphism from $\tilde{H}^{1}(\hat{\cL },\hat{\nabla })$ to ${H}^{1}_{dR}(\A \stackrel{\pi }{\rightarrow }B,\underline{{\bf C}})$.
\eop 

Although the following statement is a consequence of theorem \ref{t:1} and lemma \ref{l:4}, we will give a direct proof. 
\begin{lemma}
\label{l:5}
$\tilde{H}^{1}(\hat{\cL },\hat{\nabla })\cong {H}^{2}({\cL },{\nabla })$
\end{lemma} 
\vspace{2mm}
{\bf Proof}\hspace{2mm}
 Let $[(\hat{\cL },\hat{\nabla })]\in \tilde{H}^{1}(\hat{\cL },\hat{\nabla })$ be given. The cocycle condition states that $((\delta\hat{\cL })_\lambda ,(\delta\hat{\nabla })_\lambda )$ is equivalent with some $(\pi ^\ast {\cL }_\lambda ,\pi ^\ast {\nabla }_\lambda )$. This gives a well-defined map to $[\{ ({\cL }_\lambda ,{\nabla }_\lambda )\} ]\in {H}^{2}({\cL },{\nabla })$. Indeed, since $((\delta ^2\hat{\cL })_{\lambda\mu } ,(\delta ^2\hat{\nabla })_{\lambda\mu })$ allows a non-vanishing horizontal section, the same is true for $ ((\delta {\cL })_{\lambda\mu } ,(\delta {\nabla })_{\lambda\mu })$. Further, $[\{ ({\cL }_\lambda ,{\nabla }_\lambda )\} ]$ is independent of the choices that were made. Thus, we have constructed a homomorphism  $\tilde{H}^{1}(\hat{\cL },\hat{\nabla })\rightarrow {H}^{2}({\cL },{\nabla })$ which can be checked to be bijective. 
\eop

\begin{definition}
\label{d:85}
 Let $\tilde{H}^1_0(\hat{\cL },\hat{\nabla })$ be the subgroup of $\tilde{H}^1(\hat{\cL },\hat{\nabla })$ consisting of representatives such that $\hat{\omega }$ in eq. \Ref{eq:EQ} satisfies the integrality condition with respect to manifolds contained in a fibre.
\end{definition}
 By lemma \ref{l:35} this is equivalent with demanding that the integrality condition should be satisfied on sets $\pi ^{-1}(U_\lambda )$.
\begin{definition}
\label{d:9}
 Let $H_0(\hat{\cL },\hat{\nabla })$ be the subgroup of $\tilde{C}^{1}(\{ \pi ^{-1}(U_\lambda )\} ,\hat{\cL },\hat{\nabla })$ with representatives such that $\hat{\nabla }^2=\pi ^{\ast}{\nabla }^2$ for some connection $\nabla$ on a line bundle $\cL$ over $B$. 
\end{definition}
Using theorem \ref{t:1}, lemma \ref{l:4} and lemma \ref{l:5} the following statement is straightforward to prove. 
\begin{theorem}
\label{t:2}
\[
\begin{array}{ccccc}
{H}^{1}_{dR,0} (\A \stackrel{\pi }{\rightarrow }B,\underline{{\bf C}})& \cong & \tilde{H}^{1}_0(\hat{\cL },\hat{\nabla }) & \stackrel{\mbox{\footnotesize{monic}} }{\longleftarrow  } & {H}_0(\hat{\cL },\hat{\nabla })\nonu
\uparrow \mbox{\footnotesize{monic}}  && \uparrow \mbox{\footnotesize{monic}}  && \cong\nonu
{H}^{2}_{dR,0} (B,\underline{{\bf C}})& \cong & {H}^{2}_0({\cL },{\nabla }) & = & {H}^{2}_0({\cL },{\nabla }).
\end{array}
\]
\end{theorem}
Let us now consider the case $k=2$.
\begin{lemma}
\label{l:8}
$\tilde{H}^{2}(\hat{\cL },\hat{\nabla })\cong \tilde{H}^{2}_0(\A \stackrel{\pi }{\rightarrow }B,\underline{{\bf C}})$
\end{lemma} 
\vspace{2mm}
{\bf Proof}\hspace{2mm}
Let $[\{ (\hat{\cL }_\lambda ,\hat{\nabla }_\lambda )\} ]\in \tilde{H}^{2}(\hat{\cL },\hat{\nabla })$ be given. Let $\{ \tilde{\omega }_\lambda \} $ be the curvature 2-form of $\hat{\cL }_\lambda $. 
The cocycle relation for $\{ (\hat{\cL }_\lambda ,\hat{\nabla }_\lambda )\} $ implies that $(\delta \tilde{\omega })_{\lambda\mu }=-\pi ^\ast F_{\lambda\mu }$ for some $\{ F_{\lambda\mu }\} $. However, since $(\delta F)_{\lambda\mu\nu }=0$ we get $F_{\lambda\mu }=(\delta\omega )_{\lambda\mu }$. It leads to that $(\delta (\tilde{\omega }+\pi ^\ast \omega ))_{\lambda\mu }=0$ so the form $\hat{\omega }_\lambda := \tilde{\omega }_\lambda +\pi ^\ast \omega_\lambda $ can be written as $\hat{\omega }|_{\pi ^{-1}(U_\lambda ) }$ for some globally defined form $\hat{\omega }$ on $\A$. Thus, $\{ \tilde{\omega }_\lambda \}$ defines an element  $[\{ \tilde{\omega }_\lambda \} ]\in \tilde{H}^{2}_{0}(\{ \A \stackrel{\pi }{\rightarrow }B,\underline{{\bf C}})$. It is easy to see that $[\{ \tilde{\omega }_\lambda \} ]$ is independent of the choice of representative $\{ (\hat{\cL }_\lambda ,\hat{\nabla }_\lambda )\} $. The homomorphism $\tilde{H}^{2}(\hat{\cL },\hat{\nabla })\rightarrow \tilde{H}^{2}_0(\A \stackrel{\pi }{\rightarrow }B,\underline{{\bf C}})$ that has been constructed is clearly injective and surjective.
\eop

 As a corollary we obtain a monomorphism from ${H}^{3}_0({\cL },{\nabla })$ to $\tilde{H}^{2}(\hat{\cL },\hat{\nabla })$ as well. A direct proof of this could also have been made in a similar way as in the proof of lemma \ref{l:5}. The correspondence of theorem \ref{t:2} is:

\begin{theorem}
\label{t:3}
\[
\begin{array}{ccc}
\tilde{H}^{2}_{0}(\A \stackrel{\pi }{\rightarrow }B,\underline{{\bf C}}) & 
\cong & \tilde{H}^{2}(\hat{\cL },\hat{\nabla }) \nonu
\uparrow \mbox{\footnotesize{monic}}  && \uparrow \mbox{\footnotesize{monic}}  \nonu
\check{H}^{3}_{dR,0}(B,\underline{{\bf C}})_{(1,2)}& \cong & {H}^{3}_0({\cL },{\nabla })\nonu
\end{array}
\]
\end{theorem}

 Let us summarize some of the essentials of this section. In lemma \ref{l:2}, the de Rham cohomology on $B$ was \lq lifted\rq $ $ to an affine space $\A $. The representatives of ${H}^{2}_{dR,0} (B,\underline{{\bf C}})$ can be identified with 1-forms on $\A$. This gave a geometrical realization if considering the 1-forms as connection forms on a line bundle on $\A$. In the same way, the representatives of ${H}^{3}_{dR,0} (B,\underline{{\bf C}})$ can be identified with  2-forms on $\A$. The natural realization is curvature forms. Unfortunately though, the 2-forms are not necessary closed, while a curvature form of a line bundle is always closed due to the Bianchi identity. To solve this we use lemma \ref{l:3} which gives us closed 2-forms. The 
price we have to pay is that we get a set of 2-forms, one on each $\pi ^{-1}(U_\lambda )$. The geometrical realization of ${H}^{3}_{dR,0} (B,\underline{{\bf C}})$ is then in terms of curvatures on set of line bundles. Certainly, the same could have been done when $k=1$, but this is not necessary since a connection form doesn't have to be closed. This can be stated in a different way: The second and the third columns from the right in theorem \ref{t:1} can be used to obtain geometrical realizations for $k=1$ respective $k=2$. The fact that connection forms doesn't have to be closed makes it possible to use the right column for $k=1$. It contains to some extent simpler groups. 

When $k\geq 3$, things become more complicated. If mimicing the above approach, one is unfortunately forced to consider forms on $\A $ of at least degree 3. An alternative idea is to try to add a fourth column from the right in theorem \ref{t:1}, i.e. to find a lift of $\check{H}^{k+1}_{dR,0} (B,\underline{{\bf C}})_{(2,k-1)}$. Unfortunately, one would then come into problems which have their roots in the fact that $\{ \pi ^{-1}(U_\lambda )\} $ is not necessary a good covering. That $\{ \pi ^{-1}(U_\lambda )\} $ is not a good covering is just the typical case, as we will see in the physical examples in the next section.

\section{Anomalies and Schwinger terms}
 We will here show how the abstract constructions of the previous sections appear in a natural way from a physical model, at least for $k=1$ and $k=2$. For $k=1$ we will consider the (non-abelian) chiral anomaly in space-time and for $k=2$ the Schwinger term. For this purpose we now let $\A $ be the affine space of (external) gauge connections and $B=\A /\G$, where $\G$ is the group of base-point preserving gauge transformations (so $\A /\G$ will be a smooth manifold). The corresponding gauge group $G$ is assumed to be a compact semi-simple Lie group. Although we restrict to consider gauge theories, everything works in parallel for gravitational anomalies and Schwinger terms, \cite{EM}.

 The generating \lq functional\rq $ $ for Weyl fermions can be written as 
\[
\exp (-W(A))=\int d\psi d\bar{\psi }\exp (-\int _M \bar{\psi }\partial \!\!\! / _A^+ \psi d^{2n}x),
\]
where $W$ is the effective action and $\partial \!\!\! / _A^+$ is the positive chirality part of the anti-hermitean Dirac operator $\partial \!\!\! / _A$. The space-time $M$ is a compact, oriented and even-dimensional Riemannian spin manifold without boundary. It has been argued that the generating functional can be identified with the canonical section of the determinant line bundle $\widehat{\mbox{DET}} i\partial \!\!\! /=\mbox{det}(\mbox{ker}i\partial \!\!\! / ^+ )\otimes (\mbox{det}(\mbox{coker}i\partial \!\!\! / ^+ ))^{-1}$ over $\A$ (we assume that $\mbox{ind}\partial \!\!\! /^+=0$). The determinant line bundle has also a canonical metric, the Quillen metric. Associated with it is a natural connection $\nabla ^{ \widehat{\mbox{\footnotesize{DET}}}i\partial \!\!\! /}$ with curvature 2-form:
\eq
\label{eq:AS}
F^{ \widehat{\mbox{\footnotesize{DET}}}i\partial \!\!\! /} = -2\pi i\left( \int _M \hat{a}(M)\mbox{ch}(\ep )\right) _{[2]}, 
\eqend 
see \cite{BF}. We use $(\cdot )_{[2]}$ to denote the part of the argument that is a 2-form. $\hat{a}$ is the $a$-roof function important for gravitation and finite dimensional vector bundle $\ep $ is a the tensor product of the spinor bundle and the gauge bundle. By pushing forward all structures from $\A$ to $\A /\G$, it is seen that $F^{ \widehat{\mbox{\footnotesize{DET}}}i\partial \!\!\! /}$ can be written as $\pi ^\ast F^{\mbox{\footnotesize{DET}} {i\partial \!\!\! /}}$. The curvature 2-form $F^{\mbox{\footnotesize{DET}} {i\partial \!\!\! /}}$ can be obtained by using the family index theorem on $\A /\G$, see \cite{AS}. We have thereby argued for that $[(\widehat{\mbox{DET}} {i\partial \!\!\! /} ,\nabla ^{ \widehat{\mbox{\footnotesize{DET}}}i\partial \!\!\! /})]\in H_0(\hat{\cL },\hat{\nabla })$.

 Important is the finite anomaly ${\theta }(A;g)$ which measures the lack of gauge invariance for the effective action:
\eq
\label{eq:LG}
W(A\cdot g)=W(A)\theta (A;g)
\eqend
and the infinitesimal anomaly defined by:
\[
\hat{\omega }(A;X)=\left. \frac{d}{dt}\right| _{t=0}\theta (A;e^{tX}),
\]
where $g\in \G$ and $X\in\mbox{Lie}(\G )$. The correct interpretation in the above description is that the infinitesimal anomaly is the negative of the restriction of the connection on  $\widehat{\mbox{DET}} {i\partial \!\!\! /}$ to gauge directions, i.e. 
\eq
{\nabla }^{ \widehat{\mbox{\footnotesize{DET}}}i\partial \!\!\! /}_X\hat{s}(A)=-\hat{s}(A)\hat{\omega }(A;X).
\eqend
The minus sign originates from that the generating functional is the exponential of minus the effective action (or from the choice of Dirac operator). From eq. \Ref{eq:LG} it follows that the anomaly has to satisfy the consistency condition. Further, it is only well-defined up to changes of $W(A)$ by local functionals in $A$. It implies that the anomaly can be regarded as an element in a cohomology class. In \cite{FR} it was proven that the cohomology class of the finite anomaly is isomorphic to $\check{H}^{1}(\A /\G ,{\underline{{\bf C}}^\times })$. The free part of this states that the cohomology class of the infinitesimal anomaly is isomorphic to $\check{H}^{1}_0(\A /\G ,{\underline{{\bf C}}^\times })$, or equivalently to $ H_0(\hat{\cL },\hat{\nabla })$. Thus, the infinitesimal anomaly is described by $[(\widehat{\mbox{DET}} i\partial \!\!\! / ,\nabla ^{\widehat{\mbox{\footnotesize{DET}}} i{\partial \!\!\! /}})]\in H_0(\hat{\cL },\hat{\nabla })$ or equivalently, if disregarding torsion, by the equivalence class of line bundles on $\A /\G $ represented by $\mbox{DET} i\partial \!\!\! /$. This is the well-known result by Atiyah and Singer \cite{AS}.

 An explicit expression for the anomaly can be obtained in the following way: First use the Poincar\'{e} lemma on eq. \Ref{eq:AS} to get a local expression (up to a coboundary) of minus the connection 1-form corresponding to $\nabla ^{\widehat{\mbox{\footnotesize{DET}}} i{\partial \!\!\! /}}$. Restriction of this 1-form to gauge directions give then an expression for the anomaly. It is well known (and is easy to check) that this procedure is equivalent to the use of the descent equations. 

We have thereby shown that the cohomology groups in the previous sections are important in a physical example when $k=1$. Let us now show the correspondence for $k=2$ and Schwinger terms. We will start with an intuitive discussion before going into the mathematical details. $\A$ and $\G$ will be defined with respect to a fixed time. Our starting point will be the fact that the Schwinger term can be identified with (minus) the curvature of the vacuum line bundle over $\A$. With vacuum is meant with respect to the filled up Dirac sea. It can thus be defined as the infinite wedge product of all eigenvalues of the Hamiltonian $H_A$ which have eigenvalues less than a real number $\lambda$, the vacuum level. The ${\bf C}$-span of the vacuum defines the vacuum line bundle $\Omega _\lambda $. Clearly, it is only defined on $\hat{U}_\lambda =\{ A\in \A  | \lambda \in \!\!\!\!\! |\, \mbox{spec}(H_{A})\}$. Since the eigenvalues of $H_A$ are independent of gauge transformations, it is possible to write $\hat{U}_\lambda =\pi ^{-1} (U_\lambda )$ for some $U_\lambda \subset \A /\G$. The Schwinger term is clearly unaffected by a refinement of $\{ U_\lambda \}$ and we can therefore assume that it is an open and good covering of $\A /\G$ (if such a covering exist, see comment in the end of the section). Since the Schwinger term is coming from the curvature of the vacuum bundle, there must be some natural connection $\hat{\nabla }_\lambda $ associated with $\Omega _\lambda $. Later, when defining the vacuum line bundle rigorously, we will see how this connection is constructed. 

By applying $\hat{\delta }$ on $\{ (\Omega _\lambda ,\hat{\nabla }_\lambda )\}$ we obtain line bundles $\Omega _\lambda \otimes \Omega _\mu ^{-1} $ and connections $\hat{\nabla }_\lambda - \hat{\nabla }_\mu$ over $\pi ^{-1}(U_{\lambda \mu })$. From the intuitive definition of the vacuum bundle above it follows that $\Omega _\lambda \otimes \Omega _\mu ^{-1} $ is the ${\bf C}$-span of the finite wedge product
of all eigenvectors with eigenvalues in the interval $[\lambda ,\mu ]$. We will denote this line bundle by $\widehat{\mbox{DET}}_{\lambda \mu }$. From the gauge independence of the eigenvalues of $H_A$ it follows that $\widehat{\mbox{DET}}_{\lambda \mu }$ can be naturally pushed forward to $U_{\lambda \mu }\subset \A /\G$. Thus, if the connection $\hat{\nabla }_\lambda $ should be natural, it must be such that $((\hat{\delta }\Omega )_{\lambda \mu },(\hat{\delta }\hat{\nabla })_{\lambda \mu }  )$ is equivalent with some $(\pi ^\ast \mbox{DET}_{\lambda \mu }, \pi^\ast \nabla _{\lambda \mu })$. The reason why not $(\Omega _\lambda , \hat{\nabla }_\lambda )$ can be pushed forward is that the vacuum bundle is spanned by the infinite (and not finite) wedge product of eigenvectors. We have thereby argued for that $[\{  (\Omega _\lambda , \hat{\nabla }_\lambda )\}]\in \tilde{H}^2(\hat{\cL },\hat{ \nabla })$. In fact, as we will see later, it defines an element in the image of ${H}^3_0({\cL },{ \nabla })$ under the monomorphism in theorem \ref{t:3}. It is thus possible to use homomorphisms from previous sections. For instance, it is the image of the right monomorphism in theorem \ref{t:3} that defines the line bundles $\mbox{DET}_{\lambda \mu }$ with connections $\nabla _{\lambda \mu }$. 

 The fact that the Schwinger term can be identified with (minus) the restriction of the curvature $(\hat{\nabla }_\lambda )^2$ to gauge directions defines a homomorphism from $\tilde{H}^2(\hat{\cL },\hat{ \nabla })$ to the relevant cohomology (the one in the descent equations) of the Schwinger term.
 Non-triviality of the Schwinger term can thus be interpreted as coming from a non-zero element in ${H}^3_0({\cL },{ \nabla })$, i.e. from a set $\{ ( \mbox{DET}_{\lambda \mu }, \nabla _{\lambda \mu })\}$ such that the line bundles $\{ \mbox{DET}_{\lambda \mu }\}$ do not \lq glue together\rq $ $ due to the connections $ \{\nabla _{\lambda \mu }\}$. This is the analog of Atiyah and Singers geometrical interpretation of anomalies which in our language states that local trivial line bundles do not glue together to a global trivial bundle due to certain connections. We have thus generalized the torsion-free part of Atiyah and Singers geometrical interpretation of anomalies to the case of Schwinger terms. 

 Let us now develop the above ideas in a strict manner. A mathematical definition of the vacuum line bundle will be needed. It must also have a natural connection so that an element in $\tilde{H}^2(\hat{\cL },\hat{ \nabla })$ is obtained. 

 The Hilbert space $H$ of fermion wave functions is the space of square integrable sections of a vector bundle $\ep $ with base $M$. By abuse of notation we will use $M$ and $\ep $ to denote the physical space (no time) and the corresponding vector bundle. The Hamiltonian $H_A$ defines a decomposition $H=H_+(A,\lambda )\oplus H_-(A,\lambda )$ over $\pi ^{-1}(U_\lambda )$, where $H_-(A,\lambda )$ is the Hilbert space spanned by all eigenvalues of $H_A$ with eigenvalues less than $\lambda $. Above, we used the intuitive definition that the vacuum was the wedge product of a complete set of basis vectors in $H_-(A,\lambda )$. This definition is bad from a mathematical perspective since it is the wedge product of an infinite number of vectors. One way out of this is to \lq subtract\rq $ $ a fixed wedge product corresponding to a reference vacuum so that only a finite numbers of wedge products would be left. We will use $H_-(A_0,\lambda _0)$, for some fixed $A_0\in\pi ^{-1}({U}_{\lambda _0})$ and $\lambda _0\in {\bf R}$ to define this reference vacuum. To define the vacuum corresponding to $H_-(A,\lambda )$ we must have a way to compare it with $H_-(A_0,\lambda _0)$. We thus want to fix a connection in the trivial Hilbert space bundle over $\A$ with fibre $H(A)=H$ at $A\in\A$. The canonical identification of $H(A)$ and $H(A_0)$ given by the fact that both spaces are equal to $H$ has been used in for instance \cite{PM} to compute the Schwinger term. This gives the usual renormalization problem with Schatten classes (Hilbert-Schmidt operators in 1 space dimension). In \cite{CMM} it was shown how this can be avoided by using an alternative identification which is natural from a physical perspective. We will now review this procedure.

 Let $A(t)=f(t)A+(1-f(t))A_0$, $t\in I=[0,1]$, where $f$ is a smooth function which is 0 when $t\in [0,\epsilon ]$ and 1 when $t\in [1-\epsilon ,1]$, for some $\epsilon\in (0,\frac{1}{2})$. The Dirac equation $i\partial _t\psi =H_{A(t)}\psi$, can then be used to identify vectors in $H(A)$ with vectors in $H(A_0)$. $H_-(A,\lambda )$ is mapped to a subspace $H_-(A,\lambda )_{A_0}\subset H(A_0)$. Consider the spaces 
\begin{eqnarray*}
H_{\lambda ,+-}(A) & = & H_+(A_0,\lambda _0)\cap H_-(A,\lambda )_{A_0}\nonu
H_{\lambda ,-+}(A) & = & H_-(A_0,\lambda _0)\cap H_+(A,\lambda )_{A_0}.
\end{eqnarray*}
 We will soon show that they are finite dimensional. It gives vector bundles $H_{\lambda ,+-}$ and $H_{\lambda ,-+}$ over a subset of $\A$. By wedging together all vectors in $H_{\lambda ,+-}(A)$ respective $H_{\lambda ,-+}(A)$ we obtain line bundles $\widehat{\mbox{DET}}_{\lambda ,+-}$ and $\widehat{\mbox{DET}}_{\lambda ,-+}$. It is then natural to define $\Omega _\lambda $ as $\widehat{\mbox{DET}}_{\lambda ,+-}\otimes (\widehat{\mbox{DET}}_{\lambda ,-+})^{-1}$. This is to be thought of as the \lq subtraction\rq $ $ of the infinite wedge product of the basis vectors in $H_-(A,\lambda )$ by the corresponding product with respect to the reference space $H_-(A_0,\lambda _0)$. $\Omega _\lambda $ is however not well defined on all of $\pi ^{-1}({U}_\lambda )$ since the dimensions of $H_{\lambda ,+-}(A)$ and $H_{\lambda ,+-}(A)$ are not necessary constant. To avoid this problem we will once again redefine the vacuum bundle.

 We will now reinterpret the identification of vectors in $H(A_0)$ and $H(A)$ by using vectors $\psi$ in the Hilbert space $H\times I$ of square integrable sections of the vector bundle $\ep \times I$ over $M\times I$. The identification comes then from the $\psi$'s which satisfies $i\partial \!\!\! / _A^+\psi =0$ where $i\partial \!\!\! / _A^+=i\partial _t-H_{A(t)}$. We will choose the following boundary conditions for the partial differential equation: $\psi |_{t=0}\in H_+(A_0,\lambda _0)$ and $\psi |_{t=1}\in H_-(A,\lambda )$. By restriction to $t=0$ it is clear that $\mbox{ker}i\partial \!\!\! / ^+$ can be canonically identified with $H_{\lambda ,+-}$. Similar, $\mbox{coker}i\partial \!\!\! / ^+$ can be identified with $H_{\lambda ,-+}$. Our setting is now identical with the one used for determinant line bundles for manifolds with boundaries, see \cite{PZ}. We can therefore use the powerful results that have been achieved there. For instance, the fact that  $\mbox{ker}i\partial \!\!\! / ^+$ and $\mbox{coker}i\partial \!\!\! / ^+$ are finite dimensional. Also, we can resolve the \lq jump\rq $ $ in the dimensions by an idea given by Quillen. The result is a well-defined line bundle $\widehat{\mbox{DET}_{\lambda }}$ on $\pi ^{-1}({U}_\lambda )$. $\widehat{\mbox{DET}}_{\lambda }$ can be canonically identified with 
$\mbox{det}(\mbox{ker}i\partial \!\!\! / ^+ )\otimes (\mbox{det}(\mbox{coker}i\partial \!\!\! / ^+ ))^{-1}$ when there are no \lq jumps\rq . It also follows that $\widehat{\mbox{DET}_{\lambda }}$ has a natural metric, the Quillen metric. According to the first reference in \cite{BF}, there is a natural connection $\nabla ^{\widehat{\mbox{\footnotesize{DET}}}_\lambda }$, compatible with the Quillen metric. This motivates us to define $(\Omega _\lambda ,\hat{\nabla }_\lambda )$ as $(\widehat{\mbox{DET}}_{\lambda }, \nabla ^{\widehat{\mbox{\footnotesize{DET}}}_\lambda })$.

 We have now succeeded in finding a rigorous definition of the vacuum line bundle. Further, we have seen that it is equipped with a natural connection. Left is to show that $\{ (\Omega _\lambda ,\hat{\nabla }_\lambda )\}$ defines an element in $\tilde{H}^2_0(\hat{\cL },\hat{ \nabla })$. We will also show how an explicit expression for the Schwinger term can be obtained. 

 The expression 
\eq
\label{eq:CU}
F^{\widehat{\mbox{\footnotesize{DET}}}_\lambda }=-2\pi i \left( \int _{M\times I} \left.\hat{a}(M\times I)\mbox{ch}(\ep \times I)\right| _{\pi ^{-1}(U_\lambda )} -\frac{1}{2}\hat{\eta }_\lambda \right) _{[2]}
\eqend 
for the curvature 2-form corresponding to $\nabla ^{\widehat{\mbox{\footnotesize{DET}}}_\lambda }$ was calculated in \cite{PZ}. The only thing that we will need about $\hat{\eta }_\lambda $ is that it is defined from the boundary spectral data. It is therefore independent of gauge transformations and can thus be written as $\pi ^\ast\omega _\lambda $ for some 2-form $\omega _\lambda $ on $U_\lambda$. Since the first term on the right hand side of eq. \Ref{eq:CU} is the restriction of a form that is independent of the boundary conditions, i.e. independent of $\lambda $, we get that $F^{\widehat{\mbox{\footnotesize{DET}}}_\lambda }$ represents an element in $\tilde{H}^{2}(\A \stackrel{\pi }{\rightarrow }B,\underline{{\bf C}})$. Notice that 
\begin{eqnarray*}
&&\hat{d}(-2\pi i )\int _{M\times I} \hat{a}(M\times I)\mbox{ch}(\ep \times I) \nonu 
&& = -2\pi i \int _{M\times I} (\hat{d}+d_{M\times I}-d_{M\times I})\hat{a}(M\times I)\mbox{ch}(\ep \times I)\nonu
&& =2\pi i \int _{M\times \partial I} \hat{a}(M\times \partial I)\mbox{ch}(\ep \times \partial I) .
\end{eqnarray*}
Just as the form in \Ref{eq:AS}, the 3-form part of this can be pushed forward to $\A /\G$. The integral over any closed 3-manifold gives then an element in $2\pi i{\bf Z}$ by the index theorem. This proves that $F^{\widehat{\mbox{\footnotesize{DET}}}_\lambda }$ in fact defines an element in $\tilde{H}^{2}_0(\A \stackrel{\pi }{\rightarrow }B,\underline{{\bf C}})$ and that $[\{ (\Omega _\lambda ,\hat{\nabla }_\lambda )\}]$ is in the image of ${H}^3_0({\cL },{ \nabla })$ under the monomorphism in theorem \ref{t:3}. We can therefore use the results we arrived at in the earlier intuitive discussion. Thus, $(\Omega _\lambda ,\hat{\nabla }_\lambda )$ defines 
$({\mbox{DET}}_{\lambda\mu }, \nabla _{\lambda\mu })$ and the Schwinger terms appears since the trivial line bundles ${\mbox{DET}}_{\lambda\mu }$ do not \lq glue together\rq $ $ due to the connections $\nabla _{\lambda\mu }$. 

We will now indicate how the Schwinger term can be computed. We thus restrict eq. \Ref{eq:CU} to gauge directions. The form $\hat{\eta }_\lambda $ will then give zero contribution. We need to compute $\mbox{ch}(\ep \times I)_{[2m]}=(i/2\pi )^m\mbox{tr}{\F }^m/m!$, where $\F$ is the curvature of the bundle $\ep\times I\times\A\rightarrow M\times I\times\A$. The connection on this bundle is $A(t)$ when restricted to the the first two factors. We will only need the connection in gauge directions for the last factor. Following \cite{AS} we demand that the connection should be invariant under gauge transformations of the bundle. This gives the connection $f(t)A+(1-f(t))A_0+f(t)v=A(t)+f(t)v$, where $v$ is the ghost. It is important to remember that the gauge transformations are at a fixed time, i.e with respect to $\A$, and not $\A\times I$. Inserting this into eq. \Ref{eq:CU} and performing the integration over $I$ gives then the Schwinger term. Certainly, the computation is independent of the choice of $f$. When $A_0=0$ it gives the usual expression for the Schwinger term, which can be obtained for example by the descent equations. A non-zero $A_0$ gives the expression for the Schwinger term with a background connection. 

 There is an interesting relation between the terms in eq. \Ref{eq:CU}. The first term on the right hand side appears also for the case when manifolds without boundary are considered. It is the second term which \lq takes care\rq $ $ about the boundary conditions. From this point of view the two terms seems unrelated. Observe now that eq. \Ref{eq:CU} can be identified with eq. \Ref{eq:11}. Lemma \ref{l:3} then states that if we know the cohomology class of one of the three terms in eq. \Ref{eq:CU}, then we also know it for the other two. For example, knowing the relevant cohomology class of $ -2\pi i (\int _{M\times I} \hat{a}(M\times I)\mbox{ch}(\ep \times I))_{[2]}$, we also know it for the set $\{d\omega _\lambda \}$, where $\hat{\eta }_\lambda =\pi ^\ast \omega _\lambda $. This shows for instance that, on the level of cohomology, we never need to know the explicit expression for $\hat{\eta }_\lambda $ in this case.

In this section we have put $B=\A /\G$ and used statement from the previous sections. This means that we implicitly assume things about $\A /\G$. For instance that it admits a good covering and a partition of unity. The existence of a partition of unity was shown in \cite{CMM} by the use of a theorem by Milnor. However, that $\A /\G$ admits a good covering might not be true. That is not so crucial for this section since many of the main ideas goes through even if the covering is not good. It is straight forward to check that we then obtain the \v{C}ech cohomology $\check{H}^k(\{ U_\lambda \} ,\underline{{\bf C}}^\times )$ that depends on the covering and the corresponding cohomology $H^k(\{ U_\lambda \} ,\cL ,\sigma )$, discussed in the end of the section 2.

\thanks{\bf Acknowledgments:} I thank prof. J. Mickelsson for stimulating discussions and the International Centre for Theoretical Physics, Trieste, for their hospitality. 

\newpage
\appendix
\section*{Appendix: Proof of proposition \ref{p:2}}
 That $[F]\in H^{k+1}_{dR,0}(B,\underline{{\bf C}})$ is equivalent with that $F$ up to a (uninteresting) coboundary can be written as 
\[
F_\lambda =\sum _{\lambda _0 ,...,\lambda _{k-1}}d\rho _{\lambda _0}\wedge d\rho _{\lambda _1}\wedge ...\wedge d\rho _{\lambda _{k-1}}\wedge d\log g_{\lambda \lambda _0\lambda _1 ...\lambda _{k-1}}
\]
over $U_\lambda $, where $[\{ g_{\lambda _0...\lambda _{k}}\}]\in \check{H}^{k}(B,\underline{{\bf C}}^\times )$. We will now show that $F$ fulfills the integrality condition, i.e. that its integral over any closed $(k+1)$-dimensional manifold $S$ equals $2\pi i$ times an integer. For this purpose we assume, without loss of generality, that there exist a finite subset $\{ U_\lambda \} _{\lambda \in \Lambda_0}$ such that $V_\lambda :=U_\lambda \cap S$ are open, disjoint sets with $\mbox{dim}(\partial V_{\lambda _{i_0}}\cap ...\cap \partial V_{\lambda _{i_l}})=k-l+1$ and $S=\cup _{\Lambda\in\Lambda _0}\overline{V_\lambda }$, where the bar denotes the closure.  Then, 
\begin{eqnarray*}
\int _SF & = & \sum _{\mu _0 ,\lambda _0 ,...,\lambda _{k-1}}\int _{V_{\mu _0} }d\rho _{\lambda _0}\wedge d\rho _{\lambda _1}\wedge ...\wedge d\rho _{\lambda _{k-1}}\wedge d\log g_{\mu _0 \lambda _0\lambda _1 ...\lambda _{k-1}}\nonu
& = & \sum _{\mu _0 ,\lambda _0 ,...,\lambda _{k-1}}\int _{\partial V_{\mu _0} }\rho _{\lambda _0}\wedge d\rho _{\lambda _1}\wedge ...\wedge d\rho _{\lambda _{k-1}}\wedge d\log g_{\mu _0 \lambda _0\lambda _1 ...\lambda _{k-1}}\nonu
& = & \frac{1}{2!}\sum _{\mu _0 ,\mu _1 ,\lambda _0 ,...,\lambda _{k-1}}\int _{\partial V_{\mu _0}\cap \partial V_{\mu _1} }\rho _{\lambda _0}\wedge d\rho _{\lambda _1}\wedge ...\wedge d\rho _{\lambda _{k-1}}\nonu 
&& \wedge d(\log g_{\mu _0 \lambda _0\lambda _1 ...\lambda _{k-1}}-\log g_{\mu _1 \lambda _0\lambda _1 ...\lambda _{k-1}})\nonu
& = & \frac{1}{2!}\sum _{\mu _0 ,\mu _1 ,\lambda _1 ,...,\lambda _{k-1}}\int _{\partial V_{\mu _0}\cap \partial V_{\mu _1} }d\rho _{\lambda _1}\wedge ...\wedge d\rho _{\lambda _{k-1}}\nonu
&& \wedge d\log g_{\mu _0 \mu _1\lambda _1 ...\lambda _{k-1}},\nonu
\end{eqnarray*}
where the cocycle condition was used in the last step. The manifold $\partial V_{\mu _0}\cap \partial V_{\mu _1}$ is assumed to have orientation given by $\partial V_{\mu _0}$. This explains the relative sign in the fourth expression above. Repeating the above process until we get rid of all the partitions of unity gives:
\begin{eqnarray*}
&& \int _SF =  \frac{1}{(k+1)!}\sum _{\mu _0 ,\mu _1 ,...,\mu _{k+1}}\left.\log g_{\mu _0 \mu _1...\mu _{k}}\right| _{\partial V_{\mu _0}\cap \partial V_{\mu _1}\cap ...\cap  \partial V_{\mu _{k+1}}}\nonu 
&& =  (-1)^{k+1}\sum _{\mu _0 <\mu _1 <...<\mu _{k+1}}\sum _{i=0}^{k+1}(-1)^{i}\left.\log g_{\mu _0 ...\hat{\mu }_i...\mu _{k+1}}\right| _{\partial V_{\mu _0}\cap \partial V_{\mu _1}\cap ...\cap  \partial V_{\mu _{k+1}}}\nonu 
&& \in 2\pi i{\bf Z}.
\end{eqnarray*} 

The proof of the converse statement can be performed on each path connected component of $B$. We can thus without loss of generality assume that $\pi _0(B)=0$. Let $f_l^m$ be a  choice of representative for each element in $\pi _l(B)$, $l=0,1,...$ Then $f_l^m$ is a continuous map from the $l$-dimensional sphere to $B$. Certainly, we can assume that the sets $U_\lambda $ are star-shaped.
 Choose now  reference points $b_0\in B, b_\lambda \in U_\lambda , \forall \lambda $ and  paths $\gamma _\lambda $ from $b_0$ to $b_\lambda $. Let $b\in U_{\lambda _0...\lambda _k}$ and choose for each $i=0,1,..., k$ a  path $\triangle _{\lambda _i}^b$ from $b_{\lambda _i}$ to $b$ which depends smoothly on $b$. Then, for $j\neq i$, $[\gamma _{\lambda _i}\cup \triangle _{\lambda _i}^b \cup (\triangle _{\lambda _j}^b)^{-1}\cup (\gamma _{\lambda _j})^{-1}]=[f_1^m]\in \pi _1(B)$ for some $f_1^m=f_1^{m(i,j)}$.
 For all $i\neq j$ choose then  surfaces (2-manifolds) $\triangle _{{\lambda _i}{\lambda _j}}^b$ which have $\gamma _{\lambda _i}\cup \triangle _{\lambda _i}^b \cup (\triangle _{\lambda _j}^b)^{-1}\cup (\gamma _{\lambda _j})^{-1}\cup (f_1^m)^{-1}$ as its boundary and depends smoothly on $b$. For $l\neq i,j$, we then have $[\triangle _{{\lambda _j}{\lambda _l}}^b\cup (\triangle _{{\lambda _i}{\lambda _l}}^b)^{-1}\cup \triangle _{{\lambda _i}{\lambda _j}}^b]=[f_2^m]\in \pi _2(B)$, and can thus choose  3-manifolds $\triangle _{{\lambda _i}{\lambda _j}{\lambda _l}}^b$ with $\triangle _{{\lambda _j}{\lambda _l}}^b\cup (\triangle _{{\lambda _i}{\lambda _l}}^b)^{-1}\cup \triangle _{{\lambda _i}{\lambda _j}}^b\cup (f_2^m)^{-1}$ as boundary. $(\triangle _{{\lambda _i}{\lambda _l}}^b)^{-1}$ is the same manifold as $\triangle _{{\lambda _i}{\lambda _l}}^b$, but with the opposite orientation. Continue this procedure until $(k+1)$-manifolds $\triangle _{\lambda _0 ...\lambda _k}^b$ has been chosen. Then $\partial \triangle _{\lambda _0 ...\lambda _k}^b=(\cup _{i=0}^k{\triangle _{\lambda _0 ...\hat{\lambda }_i...\lambda _k}^b}^{(-1)^i})\cup (f_k^m)^{-1}$. From $F$ we can now construct the corresponding holonomy-free $(k-1)$-gerbe by 
\[
g_{\lambda _0 ...\lambda _k}(b)=\exp ( \int _{\triangle _{\lambda _0 ...\lambda _k}^b}F).
\]
 The cocycle property of $\{ g_{\lambda _0 ...\lambda _k}\}$ is a consequence of the integrality condition. The proof follows if we can prove that $[F]$ is the image of $[\{ g_{\lambda _0 ...\lambda _k}\}]$ under the map from $\check{H}^{k}(B,\underline{{\bf C}}^\times )$ to $H^{k+1}_{dR}(B,\underline{{\bf C}})$. Writing $F$ as $F_\lambda =\sum _{\lambda _0 ,...,\lambda _{k-1}}d\rho _{\lambda _0}\wedge d\rho _{\lambda _1}\wedge ...\wedge d\rho _{\lambda _{k-1}}\wedge \omega _{\lambda \lambda _0\lambda _1 ...\lambda _{k-1}}$ we see that we must show that $\{ \omega _{\lambda _0...\lambda _{k}}\} $ and $\{ d\log g_{\lambda _0...\lambda _{k}}\}$ represents the same element in the \v{C}ech-de Rham cohomology. To prove this we will make use of the fact that the cohomology class of $\{ g_{\lambda _0...\lambda _{k}}\} $ is independent of the choices of $b_0$, $b_\lambda $, $\gamma _\lambda $ and $\triangle _{\lambda _{i_0} ...\lambda _{i_l}}^b$. For example, since $F$ fulfills the integrality condition it is possible to replace $\triangle _{\lambda _{0} ...\lambda _{k}}^b$ with any other manifold which has the same boundary. If for $i=0,...,k$ changing $\triangle _{\lambda _{0} ...\hat{\lambda }_i...\lambda _{k}}^b$ to $\triangle _{\lambda _{0} ...\hat{\lambda }_i...\lambda _{k}}^{\prime b}$ then 
\[
\int _{\triangle _{\lambda _0 ...\lambda _k}^{\prime b}}F=\int _{\triangle _{\lambda _0 ...\lambda _k}^b}F+\sum _{i=0}^k(-1)^i\int_{W_{\lambda _{0} ...\hat{\lambda }_i...\lambda _{k}}^b} F,
\]
where $\partial W_{\lambda _{0} ...\hat{\lambda }_i...\lambda _{k}}^b=(\triangle _{\lambda _{0} ...\hat{\lambda }_i...\lambda _{k}}^b)^{-1}\cup\triangle _{\lambda _{0} ...\hat{\lambda }_i...\lambda _{k}}^{\prime b}\cup f^m_{k-1}$, where $f^m_{k-1}$ represents the same element in $\pi _{k-1}(B)$ as $(\triangle _{\lambda _{0} ...\hat{\lambda }_i...\lambda _{k}}^b)^{-1}\cup\triangle _{\lambda _{0} ...\hat{\lambda }_i...\lambda _{k}}^{\prime b}$. This only changes $\{ g_{\lambda _0...\lambda _{k}}\} $ by a coboundary. A change in $b_0$, $b_\lambda $, $\gamma _\lambda $ and $\triangle _{\lambda _{i_0} ...\lambda _{i_l}}^b$, $l\leq k-2$, does also only give a coboundary. The reason is that we can choose the $\triangle _{\lambda _{0} ...\hat{\lambda }_i...\lambda _{k}}^b$'s such that the measure of $\triangle _{\lambda _{0} ...\lambda _{k}}^b$ is unaffected by such variations (as long as the variations does not affect the point $\{ b \}$). The manifolds $b_0$, $b_\lambda $, $\gamma _\lambda $ and $\triangle _{\lambda _{i_0} ...\lambda _{i_l}}^b$, $l\leq k-2$, are thus of too low dimension to cause any changes in $\triangle _{\lambda _{0} ...\lambda _{k}}^b$'s measure. With similar methods it can also be proven that the cohomology class of $\{ g_{\lambda _0...\lambda _{k}}\} $ is independent of the choice of representatives $f^m_l$. 

 The above discussion implies that we may assume that the $b$-dependence of $g_{\lambda _0...\lambda _{k}}$ is concentrated to a small contractible neighbourhood $N\subset U_{\lambda _0...\lambda _{k}}\cap \triangle _{\lambda _{0} ...\lambda _{k}}^b$ of $\{ b\}$. Let $V_{\lambda _0},...,V_{\lambda _{k+1}}$ be open, disjoint sets with $\mbox{dim}(\partial V_{\lambda _{i_0}}\cap ...\cap \partial V_{\lambda _{i_l}})=k-l+1$, $N\subset \cup _{i=0}^k\overline{V_{\lambda _{i}}}\subset U_{\lambda _0...\lambda _{k}}\cap \triangle _{\lambda _{0} ...\lambda _{k}}^b$ and $\{ b\}\in\partial (\partial V_{\lambda _0}\cap ...\cap \partial V_{\lambda _k})$. Then, 
\begin{eqnarray*}
&& d\log g_{\lambda _0 ...\lambda _k}(b)  =  \sum _{i=0}^{k+1}d\int _{V_{\lambda _i}}F_{\lambda _i} \nonu
&& =  \sum _{i=0}^{k+1}\sum _{\mu _0,...,\mu _{k-1}}d\int _{\partial V_{\lambda _i} }\rho _{\mu _0}\wedge d\rho _{\mu _1}\wedge ...\wedge d\rho _{\mu _{k-1}}\wedge \omega _{\lambda _i \mu _0\mu _1 ...\mu _{k-1}}= ... \nonu
&& = d\int _{\partial V_{\lambda _0}\cap ...\cap  \partial V_{\lambda _{k}}}\omega _{\lambda _0 ...\lambda _{k}},
\end{eqnarray*}
where we in the last step we used the computational techniques from the first part of the proof. Notice however that $\cup _{i=0}^{k+1} \overline{V_{\lambda _i}}$ is not a closed manifold here. Equalities as $\partial V_{\lambda _i}=\sum _{j \neq i}\partial V_{\lambda _i}\cap \partial V_{\lambda _j }$ are therefore no longer true. However, the above equation anyway follows since the extra terms that appears do not depend on $\{ b\}$. Since $\omega _{\lambda _0...\lambda _{k}}$ is exact on $N$, the right hand side equals $\omega _{\lambda _0...\lambda _{k}}(b)$, which was to be proven.
\eop

\end{document}